\documentclass[fleqn]{svmult}

\usepackage{amssymb}
\usepackage{amsmath}
\usepackage{mathptmx}       
\usepackage{helvet}         
\usepackage{courier}        
\usepackage{makeidx}         
\usepackage{graphicx}        
\usepackage{multicol}        

\usepackage[usenames]{xcolor}
\usepackage{epstopdf}

\makeindex             

\begin{document}

\title*{The Relative Lyapunov Indicators: Theory and Application to Dynamical Astronomy}
\author{Zsolt S\'andor and Nicol\'as Maffione}
\institute{Zsolt S\'andor \at Computational Astrophysics Group,
Konkoly Observatory of the Hungarian Academy of Sciences, Budapest, Hungary\\
\email{zssandor@konkoly.hu}
\and Nicol\'as Maffione \at Grupo de Caos en Sistemas Hamiltonianos, Facultad de Ciencias Astron\'omicas y Geof\'isicas, \\ Universidad Nacional de la Plata, Argentina\\ \email{nmaffione@fcaglp.unlp.edu.ar}}
%
%
\maketitle

\abstract{A recently introduced chaos detection method, the Relative Lyapunov Indicator (RLI) is investigated in the cases of symplectic mappings and continuous Hamiltonian systems. It is shown that the RLI is an efficient numerical tool in determining the true nature of individual orbits, and in separating ordered and chaotic regions of the phase space of dynamical systems. A comparison between the RLI and some other variational indicators are presented, as well as the recent applications of the RLI to various problems of dynamical astronomy.}

\section{Introduction}
\label{sec:1}

One of the most important questions of investigating a dynamical system with $i>1$ degrees of freedom is to identify the ordered or chaotic behaviour of its orbits. If the dynamical system is governed by ordered orbits its time evolution is predictable. On the contrary, if the dynamical system evolves through chaotic orbits, its long-term behaviour cannot be predicted. In this paper we are considering a special class of dynamical systems called Hamiltonian systems. The phase space of a Hamiltonian-system usually contains both regions for ordered (predictable) and chaotic (unpredictable) motion, therefore the informations about the locations and extent of these regions are of high interest in investigating the evolution of such systems. A typical class of Hamiltonian systems are the planetary systems such as the Solar System, or extrasolar planetary systems. 

The ordered behaviour of an orbit or trajectory\footnote{In the case of continuous dynamical systems the \emph{trajectory} is a continuous curve in the phase space given by the points representing the time evolution of an initial state. In discrete dynamical systems the set of the discrete points representing the time evolution of the system is called as \emph{orbit}.} is strongly related to its stability. By the term \emph{stability} we mean that the trajectories are located in a bounded region of the phase space. If the region of chaotic motion is not bounded in the phase space, the trajectories could leave that domain through chaotic diffusion. In this case the chaotic trajectories become \emph{unstable}. Thus one way to perform stability investigations of dynamical systems is the detection of ordered or chaotic behaviour of their orbits. The stability of the planets or asteroids in the Solar System is an outstanding question of dynamical astronomy. The ongoing discovery of exoplanetary systems made the stability investigations of planetary systems even more important. 

In recent years there has been a growing interest in development and application of different chaos detection methods. Beside the ``traditional" tools such as the largest Lyapunov Characteristic Exponent (LCE) or Lyapunov Characteristic Number (LCN; \cite{BGGS1980}) and the frequency analysis (\cite{Laskar1993}), several new methods have been developed, which can be used to detect the ordered and chaotic nature of individual orbits, or to separate regions of ordered and chaotic motions in the phase space of a dynamical system. These methods are the Fast Lyapunov Indicator (FLI; \cite{FGL1997}; \cite{GLF2002}), the Orthogonal Fast Lyapunov Indicator (OFLI; \cite{FLFF2002}), the Mean Exponential Growth factor of Nearby Orbits (MEGNO; \cite{CS2000,CGS2003}), the Spectral Distance (SD; \cite{VCE1999}), the Smaller ALignment Index (SALI; \cite{S2001,SABV2004}), the Generalized ALingment Index (GALI; \cite{SBA2007,SBA2008}), and finally, the Relative Lyapunov Indicator (RLI; \cite{SEE2000,SESF2004}), whose analysis is the main scope of this paper. We note that all of these methods are based on the time evolution of the infinitesimally small tangent vector to the orbit, which is provided by the variational equations. Thus these chaos detection methods can be classified as \emph{variational} indicators.

In what follows, after recalling the definition of the RLI, we present its behaviour in symplectic mappings and continuous Hamiltonian systems. The efficiency of the RLI is presented by a comparative study with the already mentioned variational indicators. This paper closes with a chapter presenting the recent applications of the RLI.

\subsection{Definition of the RLI}
\label{subseq:2}
The ordered or the chaotic nature of a trajectory can be characterized most precisely by the calculation of the LCE:

\begin{equation*}
L^1(\vec{x}^*) = \lim_{t\to\infty}\frac{1}{t}\log\frac{\vec{||\xi}(t)||}{||\vec{\xi}_0||} \ ,
\end{equation*}
where $\vec{x}^* \in \mathbb{R}^n$ is the initial state of the system, or in other words, the starting point of the trajectory, and $\vec{\xi}(t)$ is the image of an initial infinitesimally small deviation vector $\xi_0$ between two nearby trajectories after time $t$. The time-evolution of $\xi$ is given by the equations of motions and their linearized equations:

\begin{equation*}
\frac{\mathrm{d}\vec{x}(t)}{\mathrm{d}t} = f[\vec{x}(t)]\ , \ \ \ \ \ \ \frac{\mathrm{d}\xi(t)}{\mathrm{d}t} = D\vec{f}[\vec{x}(t)]\xi \ ,
\end{equation*}
where $D\vec{f}[\vec{x}(t)]$ is the Jacobian matrix of the function $\vec{f}:\mathbb{R}^n\to \mathbb{R}^n$ evaluated at $\vec{x}(t)\in \mathbb{R}^n$, and $\vec{x}:\mathbb{R}\to \mathbb{R}^n$, $\xi:\mathbb{R}\to \mathbb{R}^n$ are vector-valued functions, too. In Hamiltonian systems if $L^1(\vec{x}^*) = 0$, the orbit emanating from the initial state $\vec{x}^*$ is ordered, if $L^1(\vec{x}^*) > 0$ it is chaotic. A serious disadvantage of the calculation of the LCE is that it can be obtained as a limit, thus its value can only be extrapolated, which makes the identification of weakly chaotic orbits unreliable.

In practice, one calculates only the finite-time approximation of the LCE, called the finite-time Lyapunov Indicator (LI):

\begin{equation*}
L(\vec{x},t) = \frac{1}{t}\log\frac{\vec{||\xi}(t)||}{||\vec{\xi}_0||}\ .
\end{equation*}
It is obvious that by calculating the LI for short time, the true nature of individual orbits cannot be identified. However, the basic features of the phase space of a system (the existence and approximate position of regular regions and extended chaotic domains) can be discovered very quickly by calculating a large number of LIs for short time. Let $\vec{x}$ be a vector variable taken along a line, which is going through both regular and chaotic regions of the phase space of a dynamical system. Then by fixing the integration time $t_\mathrm{int}$, one can calculate the curve $L(\vec{x}, t_\mathrm{int})$. In the case of regular regions (KAM tori, islands of stability) the curve $L(\vec{x}, t_\mathrm{int})$ varies smoothly, while in the case of an extended chaotic region it shows large fluctuations (\cite{CV1997}). However, in the case of weak chaos the fluctuations of the curve $L(\vec{x}, t_\mathrm{int})$ are not large enough to decide the true nature of the investigated region. In order to measure the fluctuations of the curve of the finite-time LI at $\vec{x}^*$, we introduce the quantity:

\begin{equation}
\Delta L(\vec{x}^*,t) = |L(\vec{x}^*+\Delta\vec{x}^*,t) - L(\vec{x}^*,t)| \ ,
\label{def}
\end{equation}
which is the difference between the finite-time LI of two neighbouring orbits, and $\Delta\vec{x}^\ast$ is the distance between the two initial condition vectors. This quantity has been introduced and called RLI in \cite{SEE2000,SESF2004}. Definition \eqref{def} contains $\Delta\vec{x}^\ast$ as a free parameter, which should be chosen small enough to reflect the local properties of the phase space. In our numerical investigations we have experienced that the arbitrary choice of $\Delta\vec{x}^\ast$ in a quite large interval $||\Delta\vec{x}^\ast|| \in [10^{-14}, 10^{-7}]$ does not modify essentially the behaviour of the RLI as a function of the time. For ordered orbits the RLI shows linear dependence on $||\Delta\vec{x}^\ast||$, while for chaotic orbits the RLI practically is invariant with respect to the choice of $||\Delta\vec{x}^\ast||$. 

Although there is not developed a strict mathematical theory describing the time behaviour of the RLI so far, the results of numerical simulations clearly show its power in separating ordered and chaotic orbits. An intuitive explanation could be that in the case of ordered orbits the time evolution of the two LI curves ($L(\vec{x}^*,t)$ and $L(\vec{x}^*+\Delta\vec{x}^*,t)$) practically cannot be distinguished meaning that they converge wiht the same (or very similar) rate to the LCE = 0 limit. On the other hand, the convergence rate of the LI of two close chaotic orbits (separated in the phase space by the vector $\Delta\vec{x}^\ast$) could be very different, which is reflected in the time evolution of the RLI. In the next sections of the paper we shall investigate through extensive numerical experiments the completely different behaviour of the RLI as a function of time in the cases of ordered and chaotic orbits, which makes it a suitable tool of chaos detection.

\subsection{Properties of the RLI in chaos detection}
\label{subseq:3}

In order to eliminate the high frequency fluctuations of the curve $\Delta L(\vec{x},t)$ for a fixed $\vec{x}\in\mathbb{R}^n$, we suggested the following smoothing

\begin{equation*}
\langle\Delta L(\vec{x})\rangle(t) = \frac{1}{t}\sum_{i=1}^{[t/\Delta t]} \Delta L(\vec{x},i\cdot\Delta t)\ ,
\end{equation*}
where $\Delta t$ is the stepsize. In numerical experiments we always use the above smoothed value of the RLI. 

The different behaviour of the RLI for ordered and chaotic orbits are first presented for discrete Hamiltonian systems, such as the following 2D : 
\begin{equation}
\left\{
\begin{array}{lll}
x_1^{\prime}&=&x_1+x_2\\
x_2^{\prime}&=&x_2-\nu\cdot\sin(x_1+x_2) \mod(2\pi)\ ,
\end{array}
\right.
\label{2dmap}
\end{equation}

\noindent
and 4D symplectic mapping:
\begin{equation}
\left\{
\begin{array}{lll}
x_1^{\prime}&=&x_1+x_2\\
x_2^{\prime}&=&x_2-\nu\cdot\sin(x_1+x_2)-\mu\cdot[1-\cos(x_1+x_2+x_3+x_4)]\\
x_3^{\prime}&=&x_3+x_4\\
x_4^{\prime}&=&x_4-\kappa\cdot\sin(x_3+x_4)-\mu\cdot[1-\cos(x_1+x_2+x_3+x_4)] \mod(2\pi)\ ,
\end{array}
\right.
\label{4dmap}
\end{equation}
where $\nu$ and $\kappa$ are the non-linearity parameters, and $\mu$ is the coupling parameter of the 4D mapping.
 
\begin{figure}[t]
\includegraphics[scale=.45]{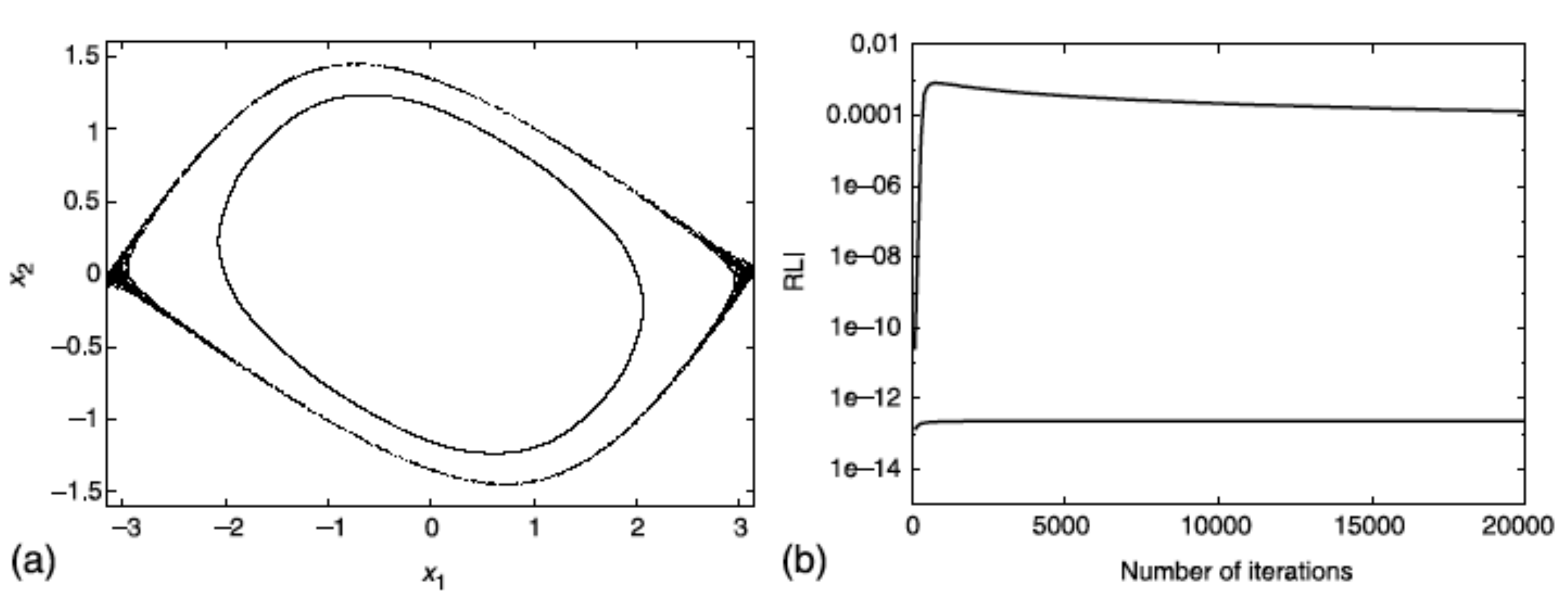}
\caption{(a) (left): The phase plot of an ordered and a chaotic orbit in the mapping \eqref{2dmap}; (b) (right):
The behaviour of the RLI as the function of time for a chaotic orbit (upper curve) and for an ordered
orbit (lower curve).}
\label{fig1.1}      
\end{figure}

\begin{figure}[t]
\includegraphics[scale=.46]{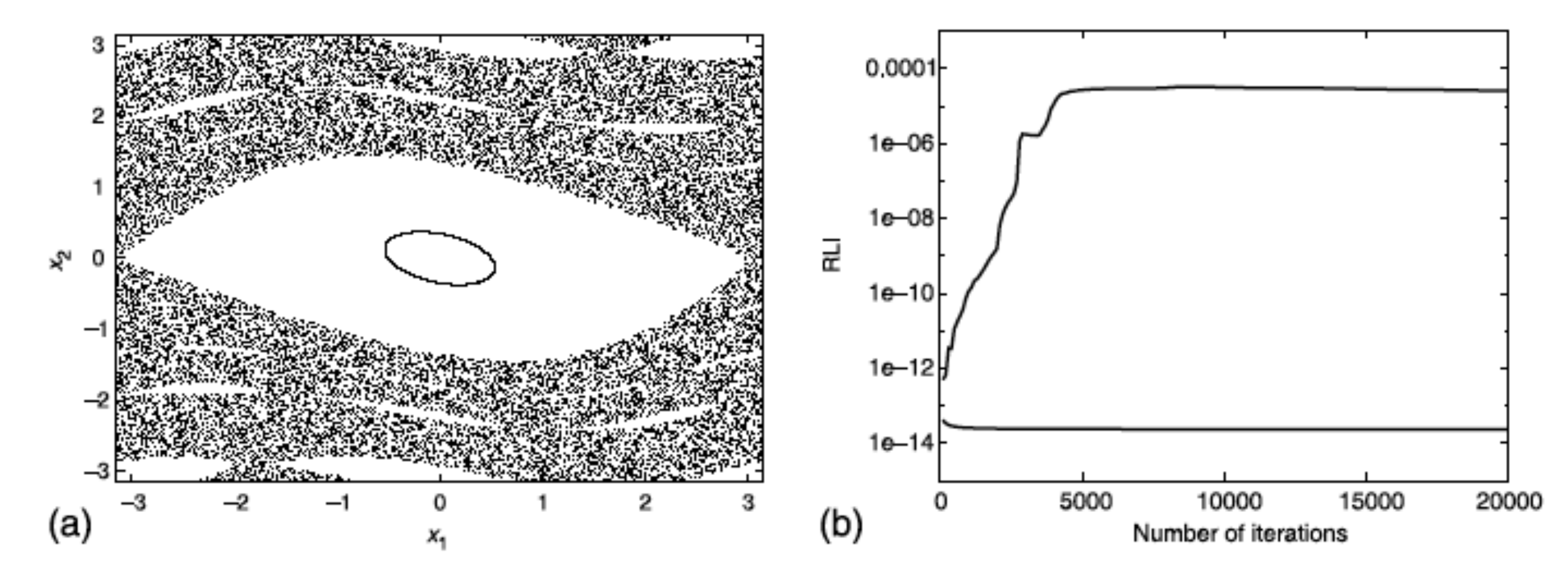}
\caption{(a) (left): The phase plot of an ordered and a chaotic orbit in the mapping \eqref{4dmap}; (b) (right):
The behaviour of the RLI as the function of time for a chaotic orbit (upper curve) and for an ordered
orbit (lower curve).}
\label{fig1.2}       
\end{figure}

In the case of the 2D symplectic mapping \eqref{2dmap} the initial conditions of the ordered orbit are $x_1=2$, $x_2=0$, while the initial conditions of the chaotic orbit are $x_1=3$, $x_2=0$. In both cases $\nu=0.5$. The phase plots of these orbits are shown in Fig. \ref{fig1.1}(a) and the corresponding time behaviour of the RLI is displayed in Fig. \ref{fig1.1}(b). In the case of the 4D mapping \eqref{4dmap} the following initial conditions are used: $x_1=0.5$, $x_2=0$, $x_3=0.5$, $x_4=0$ for the ordered, and  $x_1=3$, $x_2=0$, $x_3=0.5$, $x_4=0$ for the chaotic orbit. In both cases the parameters are $\nu=0.5$, $\kappa=0.1$ and $\mu=0.001$. The projections of these orbits onto the $x_1-x_2$ plane are shown in Fig. \ref{fig1.2}(a). The behaviour of the RLI as the functions of time of the ordered and chaotic orbits are plotted in Fig. \ref{fig1.2}(b). Studying Figs. \ref{fig1.1}(b) and \ref{fig1.2}(b) one can see that the RLI for an ordered orbit is almost constant. The RLI of a chaotic orbit grows very rapidly, and after reaching a maximum value it decreases very slowly.

The maximum value of the RLI of a chaotic orbit is much higher (in the examples shown by 9--10 orders of magnitude) than the almost constant value of the RLI of an ordered orbit. It can be seen that by using the RLI, the ordered or chaotic nature of orbits can be identified after a few hundred iterations of the investigated mapping. 
 
\begin{figure}[t]
\includegraphics[scale=.46]{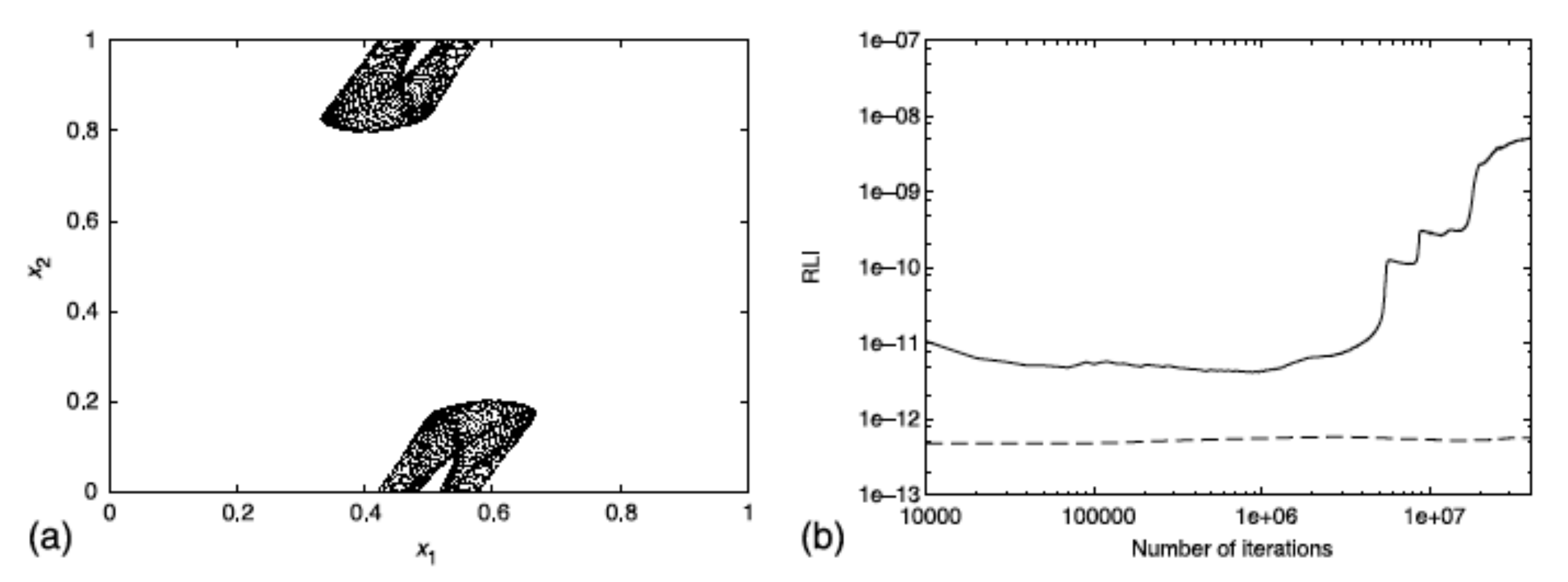}
\caption{(a) (left): The phase plot of a weakly chaotic orbit in the mapping \eqref{4dmapsticky}; (b) (right): The behaviour of the RLI as the function of time for the weakly chaotic orbit seen in the left panel (upper curve) and for an ordered orbit (lower curve).}
\label{fig1.3}       
\end{figure}

A crucial test for a chaos detection method is whether it separates the weakly chaotic orbits (also called ``sticky" orbits) from the ordered orbits. In order to demonstrate this property of the RLI we used the 4D symplectic mapping and initial conditions for an ordered and a weakly chaotic orbit as \cite{VCE1999,S2001}:  

\begin{equation}
\left\{
\begin{array}{lll}
x_1^{\prime}&=&x_1+x_2^{\prime}\\
x_2^{\prime}&=&x_2 + (K/2\pi)\sin(2\pi x_1) - (\beta/\pi)\sin[2\pi(x_3-x_1)]\\
x_3^{\prime}&=&x_3+x_4^{\prime}\\
x_4^{\prime}&=&x_4 + (K/2\pi)\sin(2\pi x_3) - (\beta/\pi)\sin[2\pi(x_1-x_3)]\mod(1)\ ,
\end{array}
\right.
\label{4dmapsticky}
\end{equation}
where $K$ is the non-linearity and $\beta$ is the coupling parameter. According to \cite{VCE1999} the orbit with the parameters $K=3$, $\beta = 0.1$ and with initial conditions $x_1 = 0.55$, $x_2 = 0.1$, $x_3 = 0.62$, $x_4 = 0.2$ is ordered, while the orbit with the same initial conditions and parameter $K$, but with $\beta = 0.3051$ is slightly chaotic tending to a very small LCE. The projection of the weakly chaotic orbit on the $x_1-x_2$ plane is shown in Fig. \ref{fig1.3}(a), and it seems to be an ordered orbit on a torus. Using the RLI (Fig. \ref{fig1.3}(b)) one can see that the chaotic nature of this orbit can be detected after about $N \sim 5 \times 10^6$ iterations.

\begin{figure}[t]
\includegraphics[scale=.46]{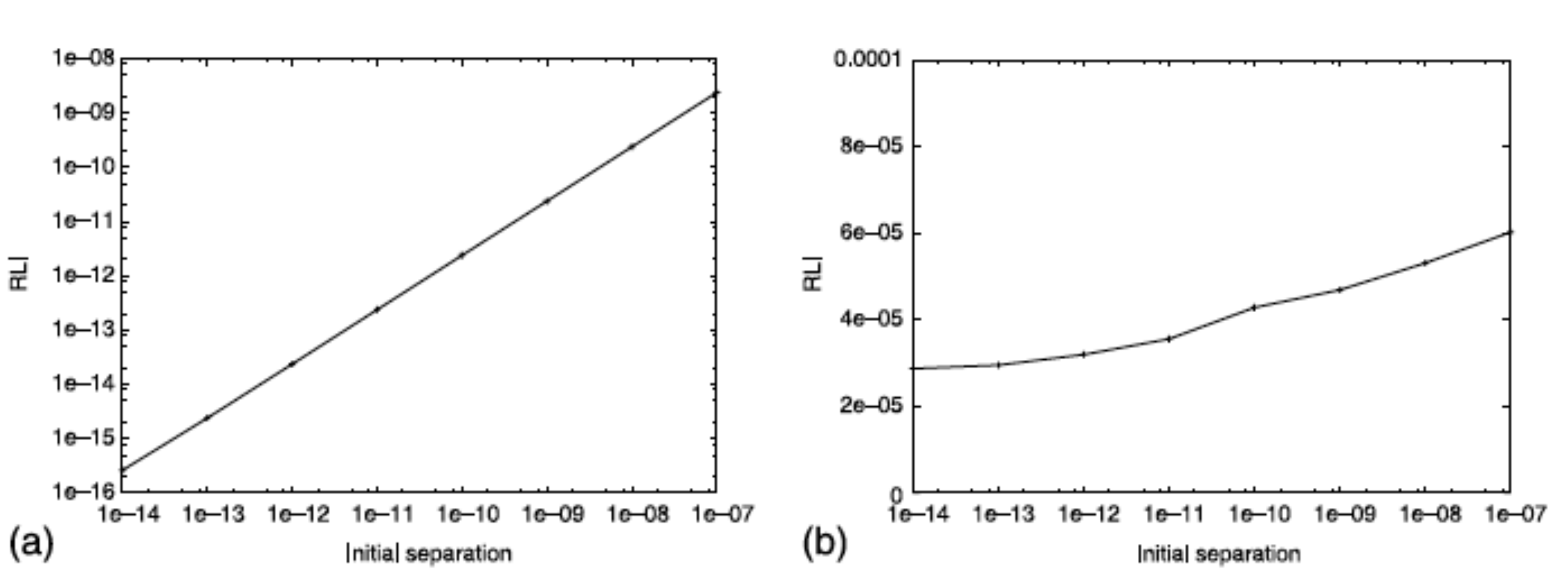}
\caption{(a) (left): The linear dependence of the final value of the RLI on the initial separation for an ordered orbit; (b) (right): The final value of the RLI versus the initial separation for a chaotic orbit.}
\label{fig1.4}       
\end{figure}

Finally, we should discuss the role of the initial separation between the two orbits, which is one of the free parameters of the RLI (the other one is the lenght of the time needed to calculate the RLI, as will be mentioned later on). In what follows, we give an evidence that $||\Delta\vec{x}^*||$ can be chosen arbitrarily from a quite large interval $[10^{-14}, 10^{-7}]$. The smallest value of this interval is due to the finite representation of numbers by computers. On the other hand, the largest value of the above interval should also be small enough in order to the RLI reflect the local property of the phase space around the orbit under study. 

Fig. (\ref{fig1.4}) shows the dependence of the RLI (obtained after a fixed number of iterations, which in this particular case was $2\times 10^4$), on $||\Delta\vec{x}^*||$ for the 4D ordered orbit of mapping \eqref{4dmap}, shown in Fig. \ref{fig1.2}(a). In Fig. \ref{fig1.4} both the horizontal and vertical axes are scaled logarithmically. Studying Fig. \ref{fig1.4}(a) one can see that for the ordered orbit the RLI changes linearly with respect to $||\Delta\vec{x}^*||$. In Fig. \ref{fig1.4}(b) we display the dependence of the RLI (after $2\times10^4$ iterations) on $||\Delta\vec{x}^*||$ for the chaotic 4D orbit shown in Fig. \ref{fig1.2}(a). One can see that in the chaotic case the final value of the RLI practically does not depend on the choice of $||\Delta\vec{x}^*||$. Thus in the above sense the RLI is invariant with respect to the choice of the initial separation. This invariance property can be explained by the fact that the RLI of a chaotic orbit practically measures the average width of the oscillation of the LI curve (as a function of time), which does not depend heavily on the choice of the initial separation. 

\section{A short comparison of the RLI to other methods of chaos detection}
\label{sec:3}


In this section we compare the performances of the RLI with the variational indicators mentioned in Sect. \ref{sec:1}: the LI, the FLI and the OFLI, the MEGNO, the SD, the SALI and the GALI.

In Sect. \ref{sec:3.1} we analyze the dependence of the RLI on its free parameters and in the following one we compare the typical behaviours of the RLI with other techniques in the well--known H\'enon--Heiles model \cite{HH1964} (hereinafter HH model). In Sect. \ref{sec:3.3} we apply the RLI and other indicators to study dynamical systems of different complexity: the 4D symplectic mapping \eqref{4dmap} presented in Sect. \ref{subseq:3} and a rather complex 3D potential resembling a Navarro, Frenk and White triaxial halo (hereinafter NFW model; \cite{VWHS2008}). We compare the phase space portraits given by the RLI and the other methods to decide whether the results are comparable. In Sect. \ref{sec:3.4} we briefly discuss the dependence of the RLI on the computing times.

The Bulirsch--Stoer integrator is used throughout this section. 

\subsection{The dependence of the RLI on the free parameters}
\label{sec:3.1}

The RLI has two free parameters: a) the initial separation $(\Delta\vec{x}^\ast)$ between the basis orbit and its ``shadow'' (Sect. \ref{subseq:2}) and b) its threshold (the threshold is a value that separates chaotic from regular motion and it is related with the lenght of the time needed to calculate the RLI). These free parameters are ``user--choice'' quantities. Thus, it is of interest to study the dependence of the RLI on both of them. 

The following experiments are undertaken on the HH model:

\begin{equation*}
\mathcal{H}=\frac{1}{2}\left(p_x^2+p_y^2\right)+\frac{1}{2}\left(x^2+y^2\right)+x^2\cdot y-\frac{1}{3}y^3\ ,
\end{equation*}
where $\mathcal{H}$ is the Hamiltonian and $x,y,p_x,p_y$ are the usual phase space variables. 

\subsubsection{The initial separation parameter}
\label{sec:3.1.1}

The dependence of the RLI on the initial separation parameter is strongly related to the type of orbit under study (see Sect. \ref{subseq:2}). Therefore, we take four different types of orbits with initial conditions located on the line defined by $x=p_y=0$ and $y\in [-0.1:0.1]$ and the energy surface $E=0.118$, namely, a regular orbit close to a stable periodic orbit (r-sp); a quasiperiodic orbit (r-qp); a regular orbit close to an unstable periodic orbit (r-up); and a chaotic orbit inside a stochastic layer (c-sl). The initial conditions are taken from \cite{CGS2003}. The integration time is $10^{4}$ units of time (hereinafter u.t.), which is enough time to provide a reliable characterization of the orbits and the stepsize of the numerical integration is $0.01$. We note that these values have been used in the following numerical experiments, too.

\begin{figure}[t]
\sidecaption[t]
\begin{tabular}{cc}
\includegraphics[scale=.46]{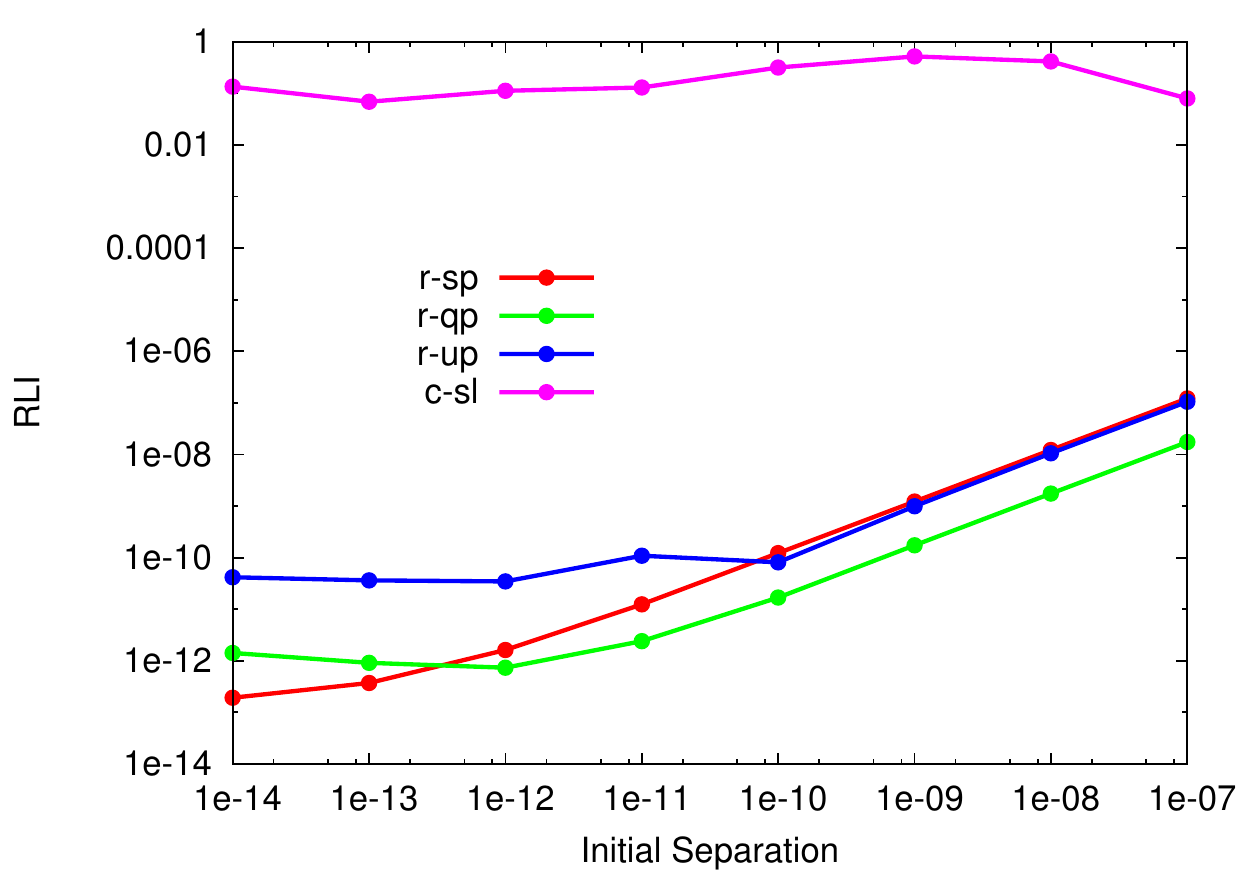}&
\includegraphics[scale=.46]{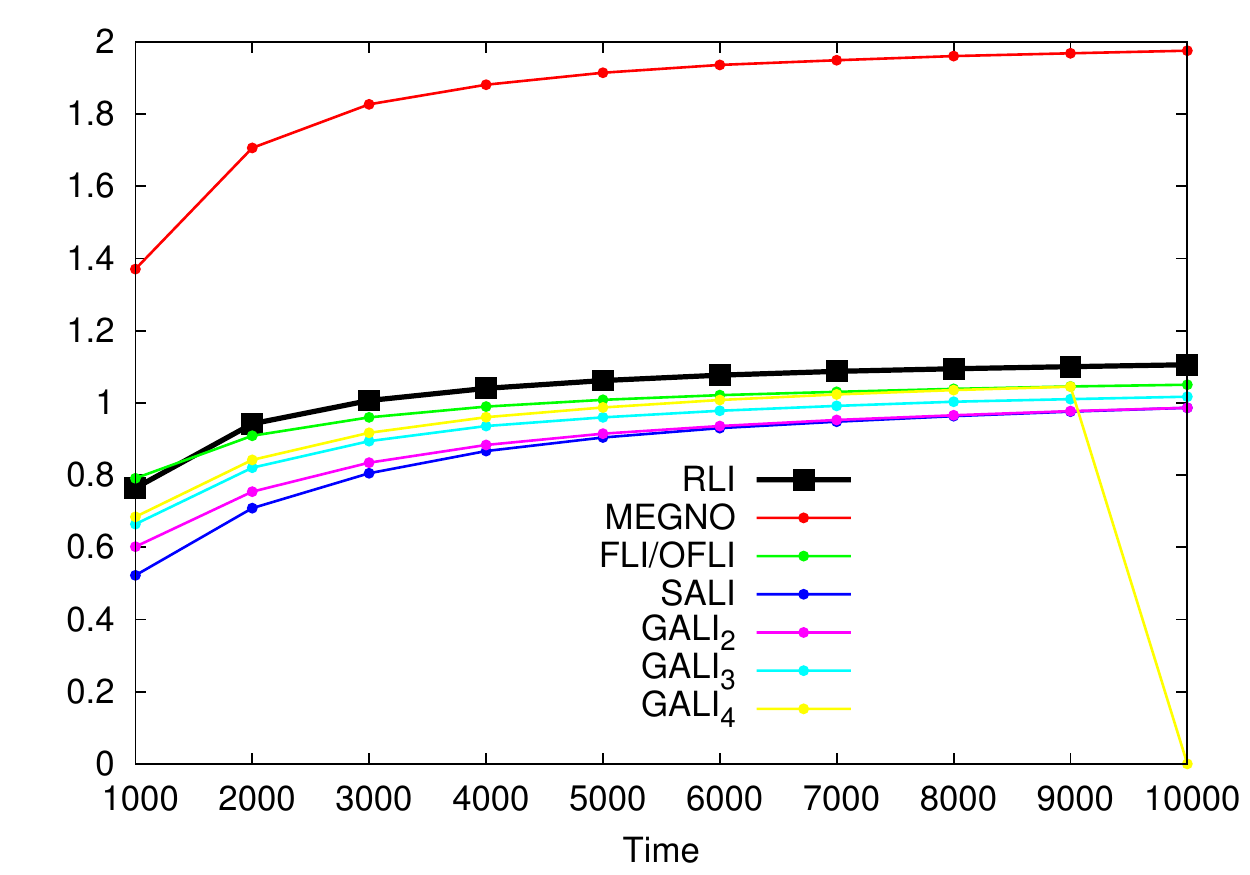}
\end{tabular}
\caption{(a) (left): The RLI values as a function of the initial separation parameter for the orbits ``r-sp'', ``r-qp'', ``r-up'' and ``c-sl''; (b) (right): The normalized approximation rates for several chaos indicators, including the RLI (see text for further details).} 
\label{rliinisep}
\end{figure}

In Fig. \ref{rliinisep}(a), we present the final values (i.e. the values of the indicator at the end of the integration time) of the RLI as a function of the initial separation parameter\footnote{The values for the parameter have been taken from the interval suggested in Sect. \ref{subseq:3}.} for the orbits introduced earlier. We show that the initial separation parameter does not significantly affect the RLI when we apply the indicator to the chaotic orbit ``c-sl'', but it does when we apply it to the regular orbits ``r-sp'', ``r-qp'' and ``r-up'' (and confirming the results shown in Sect. \ref{subseq:3}). This dependence of the RLI on its free parameter has severe implications in the selection of the threshold. For instance, if we start the computation with an initial separation of $10^{-14}$, the relation shown in Fig. \ref{rliinisep}(a) will indicate that a good candidate for the threshold to distinguish between the chaotic orbit ``c-sl'' (RLI$\sim0.1$) and the regular orbit ``r-sp'' (RLI$\sim10^{-13.5}$) can be $10^{-12}$. Then, the orbits with values of the RLI higher than $10^{-12}$ will be classified as chaotic orbits. However, this choice of the threshold leads to a misclassification of the regular orbits ``r-up'' (RLI$\sim10^{-10}$) and ``r-qp'' (RLI$\sim10^{-12}$). Furthermore, since the correspondence between the RLI and the initial separation parameter for the regular orbits of the sample tends to be linear (see Sect. \ref{subseq:3}), the threshold $10^{-12}$ does not work at all for an initial separation greater than $10^{-11}$.

In order to determine a reliable threshold for the RLI, the relationship with the initial separation parameter should be done by computing the indicator for a group of orbits known to be regular but with some level of instability (e.g. regular orbits close to a hyperbolic object such as an unstable periodic orbit, see \cite{MDCG2011}). 

\subsubsection{The threshold}
\label{sec:3.1.2}

In this section we investigate how the RLI and other indicators (listed above) depend on their thresholds. For the following experiment in the HH model we have adopted a sample of $125751$ initial conditions in the region defined by $x=0$, $y\in[-0.1:0.1]$, $p_y\in[-0.05:0.05]$ and on the energy surface defined by $E=0.118$. The thresholds of the LI, the RLI, the MEGNO, the SALI, the FLI/OFLI and the GALIs are shown in Table \ref{tablethreshold}, where $t$ is the time (see \cite{DMCG2012}). From here, the initial separation parameter will be $10^{-12}$, unless stated otherwise. The threshold used for the RLI has been computed following the remarks discussed in the previous section. 

\begin{table}\centering
\caption{Thresholds for several indicators, including the RLI}
\label{tablethreshold}
\begin{tabular}{p{2cm}p{2cm}}
\hline\noalign{\smallskip}
Indicator & Threshold \\ 
\noalign{\smallskip}\svhline\noalign{\smallskip}
LI & $ln(t)/t$ \\
RLI & $10^{-10}$ \\
MEGNO & $2$ \\
SALI & $10^{-4}$ \\
FLI/OFLI & $t$ \\
GALI$_2$ & $t^{-1}$ \\
GALI$_3$ & $t^{-2}$ \\
GALI$_4$ & $t^{-4}$ \\
\noalign{\smallskip}\hline\noalign{\smallskip}
\end{tabular}
\end{table}

To proceed with the experiment we define the approximation rate as the rate of convergence with a final percentage of chaotic orbits. This rate will show a combination of the reliability of the indicator and the accuracy of the selected threshold if the final percentage of chaotic orbits approaches the ``true percentage'' of chaotic orbits in the system. Therefore, as we require a reliable final percentage of chaotic orbits, we consider the percentage of chaotic orbits given by the LI by $10^4$ u.t.: $\sim 39.92$\%, the ``true percentage'' of chaotic orbits in the system. Both the overwhelming number of papers claiming the reliability of the LI as a chaos indicator and the experimental evidence showing that $10^4$ u.t. seems to be a reliable convergent time for all the indicators in the experiment (see, for instance, Sect. \ref{sec:3.2}, Fig. \ref{hh}) supporting this statement. Thus, we normalize the time evolution of the percentage of chaotic orbits given by the methods with this ``true percentage''. Hence, the values of the normalized approximation rates higher than 1 show percentages of chaotic orbits higher than the ``true percentage''. 

We test the reliability of the thresholds given in Table \ref{tablethreshold} according to the above mentioned rates. The results are indicated in Fig. \ref{rliinisep}(b). The convergence towards a constant rate of the RLI and the FLI/OFLI is faster than that of the other indicators of the sample. Despite this rapid tendency towards a constant value for both indicators, the final percentages of chaotic orbits given by the RLI and the FLI/OFLI are higher than that of the LI, which means that the values of the rates are above 1. Nevertheless, this slight difference in the final percentages of chaotic orbits can always be fixed with a small adjustment of the corresponding thresholds. Since the results for the MEGNO show substantial disagreement between the percentages of the chaotic orbits given by the indicator and the LI, a significant empirical adjustment of the MEGNO's threshold should be made to avoid an overestimation in the number of chaotic orbits. The final percentages given by the SALI and the GALIs are in perfect agreement with the ``true percentage''. However, their tendency towards a stable percentage of chaotic orbits is slower than the one showed by the RLI or the FLI/OFLI.

Thus, among the above mentioned CIs, the RLI and the FLI/OFLI show the best approximation rates, i.e. the best combination of the reliability of the indicators and the accuracy of their thresholds. For further details on the experiment, refer to \cite{DMCG2012}.

By $10^4$ u.t. the threshold taken for the GALI$_4$ reaches the computer's precision ($10^{-16}$) and thus, every chaotic orbit lies beyond such precision. Therefore, the last yellow point in Fig. \ref{rliinisep}(b) falls apart from the tendency.

Now that we have finished studying the importance of a wise selection of the free parameters of the RLI, we calibrated the indicator following the suggestions mentioned here and continue comparing its performance with other indicators.

\subsection{Expected behaviour of the indicators in the HH model}
\label{sec:3.2} 

In this experiment, done in the HH model, our goal is to compare the typical behaviours of the RLI to the other techniques and show its advantages and disadvantages.  

We take the orbits of Sect. \ref{sec:3.1.1} and a chaotic orbit inside the chaotic sea (c-cs) with initial conditions on the line defined by $x=p_y=0$ and $y\in [-0.1:0.1]$ and the energy surface $E=0.118$ (the initial conditions are taken from \cite{CGS2003}). The final integration time is $10^4$ u.t. and the stepsize is $0.01$. 

\begin{figure}[t]
\sidecaption[t]
\includegraphics[scale=.46]{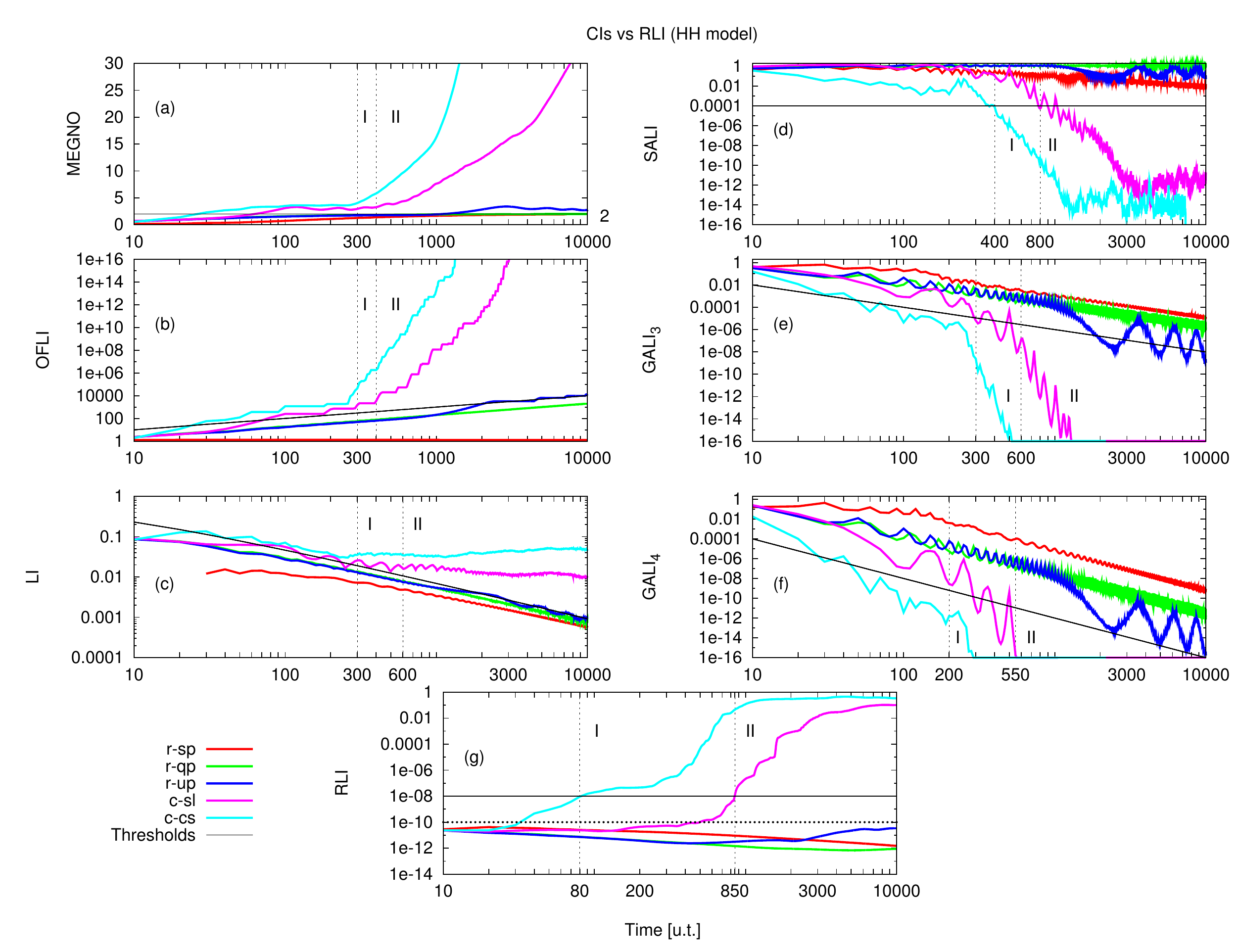}
\caption{Behaviours of (a) the MEGNO, (b) the OFLI, (c) the LI, (d) the SALI, (e) the GALI$_3$, (f) the GALI$_4$ and (g) the RLI for orbits ``r-sp'', ``r-qp'', ``r-up'', ``c-sl'' and ``c-cs''. The thresholds as well as the lines ``I'' and ``II'' are included (see text for further details).}
\label{hh}
\end{figure}

Fig. \ref{hh} shows the time evolution curves of several indicators for the five different types of orbits introduced at the beginning of the section. Some of the main features of a chaos indicator are the speed of convergence and the resolving power. The former is the time it takes to distinguish between chaotic and regular motion. In order to visualize this quantity, in Fig. \ref{hh} we introduce the vertical lines ``I'' and ``II''. The first one shows the time after which the orbit ``c-cs'' is clearly identified as a chaotic orbit with the indicator and the second one plays the same role as ``I'' for the orbit ``c-sl''. 

On panel (a) we present the typical behaviours of the MEGNO (see \cite{CS2000}). The values of the indicator for the three regular orbits tend towards the theoretical asymptotic threshold, $2$, in different ways (see the right bottom of panel a). The values for the chaotic orbits increase linearly with time. At $\sim$300 u.t. (see line ``I''), the separation between orbit ``c-cs'' and the threshold line is already significant. Hence, orbit ``c-cs'' is clearly identified as a chaotic orbit then. Only 100 u.t. later (line ``II'') the same happens with orbit ``c-sl''. 

On panel (b) we show the time evolution curves of the OFLI (\cite{FLFF2002}). The values of the indicator for two of the regular orbits increase linearly with time (see the threshold and its expression in Table \ref{tablethreshold}) while the values for the chaotic orbits increase exponentially fast with time. This distinction between both tendencies can be made at the same times that have been shown by the MEGNO. Besides, orbit ``r-sp'' has an almost constant value because it is very close to a stable periodic orbit. 

On panel (c) we present the LI (see e.g. \cite{S2010}). The values of the indicator for the three regular orbits decrease with time (see the threshold and its expresion in Table \ref{tablethreshold}) while the values for the chaotic orbits tend towards a constant value which depends on the chaoticity of the orbit. The distinction between the regular orbits and the chaotic orbit ``c-sl'' is made by the LI later than by the previous indicators: the orbit leaves the linear tendency of the threshold around $\sim$600 u.t. (line ``II''). 

On panel (d) we present the time evolution curves of the SALI (\cite{S2001}). The values of the indicator for the regular orbits ``r-qp'' and ``r-up'' oscillate within the interval (0,2), while the orbit ``r-sp'' tends towards 0 following a power law behaviour. The chaotic orbits decrease exponentially fast with time. The time needed for the SALI to clearly identify the chaotic orbits is the time used by the chaotic orbits to reach the threshold (see its value in Table \ref{tablethreshold} and lines ``I'' and ``II'' in the figure to locate the times). The indicator delays making this distinction (in fact, it does so later than the LI) because for smaller values of the integration time, the chaotic orbits decrease with a power law as the regular orbit ``r-sp'' does.

On panels (e) and (f) we show the time evolution curves of the GALI$_3$ and the GALI$_4$, respectively (the GALI$_2$ and the SALI have almost identical behaviours and the former is not included). Their theoretical thresholds (Table \ref{tablethreshold}; see \cite{CB2006,SBA2007,SBA2008,MSA2012} for further details) yield good estimations of the time needed for the indicators to distinguish the chaotic orbits from the regular orbits. The GALI$_3$ makes this distinction in the same time as the LI did (see lines ``I'' and ``II''). Once again, the reason for this delay is that the GALI$_3$ decreases with a power law for regular orbits as well as for chaotic orbits at the beginning of the integration interval. Nevertheless, the higher the order of the GALI, the faster its tendency towards 0 for the chaotic orbits. Then, the GALI$_4$ has registered the best time so far to distinguish the chaotic orbit ``c-cs'': $\sim$200 u.t.

On panel (g) we present the time evolution curves of the RLI (Sect. \ref{subseq:3}). The values of the indicator for the three regular orbits are in the interval $(10^{-12},10^{-10})$ according to the initial separation of $10^{-12}$ (see Fig. \ref{rliinisep}(a), in Sect. \ref{sec:3.1.1}). Thus, as the value $10^{-10}$ (depicted with a dotted line in the figure) that have been selected in Sect. \ref{sec:3.1.2} is in the limit of the interval, it is not reliable as a threshold any more. Therefore, we selected the value $10^{-8}$ for the threshold. The characterization of the regular orbits does not clearly differentiate among them as the MEGNO, the OFLI or the SALI. The values for the chaotic orbits increase with time until they reach a constant value. On the one hand, orbit ``c-cs'' is clearly identified as a chaotic orbit by $\sim$80 u.t. (line ``I'') when the orbit reaches the threshold. This is the fastest characterization of the chaotic orbit ``c-cs''. On the other hand, the RLI identifies orbit ``c-sl'' as a chaotic orbit around $\sim$850 u.t. (line ``II''), which is the slowest characterization.  

All the indicators delay in making a reliable characterization of the chaotic orbit ``c-sl'', which shows that the chaotic orbit ``c-cs'' has a larger LI than orbit ``c-sl''.

Finally, the characterization of the five representative orbits made by the RLI as well as its speed of convergence is similar to the other techniques. Thus, the RLI is most welcome to the group of fast variational indicators.  

\subsection{Performances of the indicators under different scenarios}
\label{sec:3.3}

We have seen in the previous section how similar are the performances of the RLI and the other fast indicators in the rather simple HH model. Here we will focus on experiments in scenarios that are different from the HH model to determine whether the RLI is a reliable technique for studying different or more complex systems than the HH model. 

\subsubsection{The 4D symplectic mapping}
\label{sec:3.3.1}

The time evolution curves of the indicators (used in Sect. \ref{sec:3.2}) are not efficient to analyze a large number of orbits. The appropriate way to gather information in these cases is in terms of the final values of the methods. Thus, let us now turn our attention to the study of the resolving power of the techniques using their final values. 

The following study will be conducted in the 4D mapping \eqref{4dmap} presented in Sect. \ref{subseq:3} by adopting different samples of initial conditions and $10^5$ iterations. The version of the MEGNO considered here is the MEGNO(2,0), whose threshold value is $0.5$ (see \cite{CGS2003}).

\begin{figure}[t]
\sidecaption[t]
\begin{tabular}{ccc}
\hspace{-8mm}\includegraphics[scale=.40]{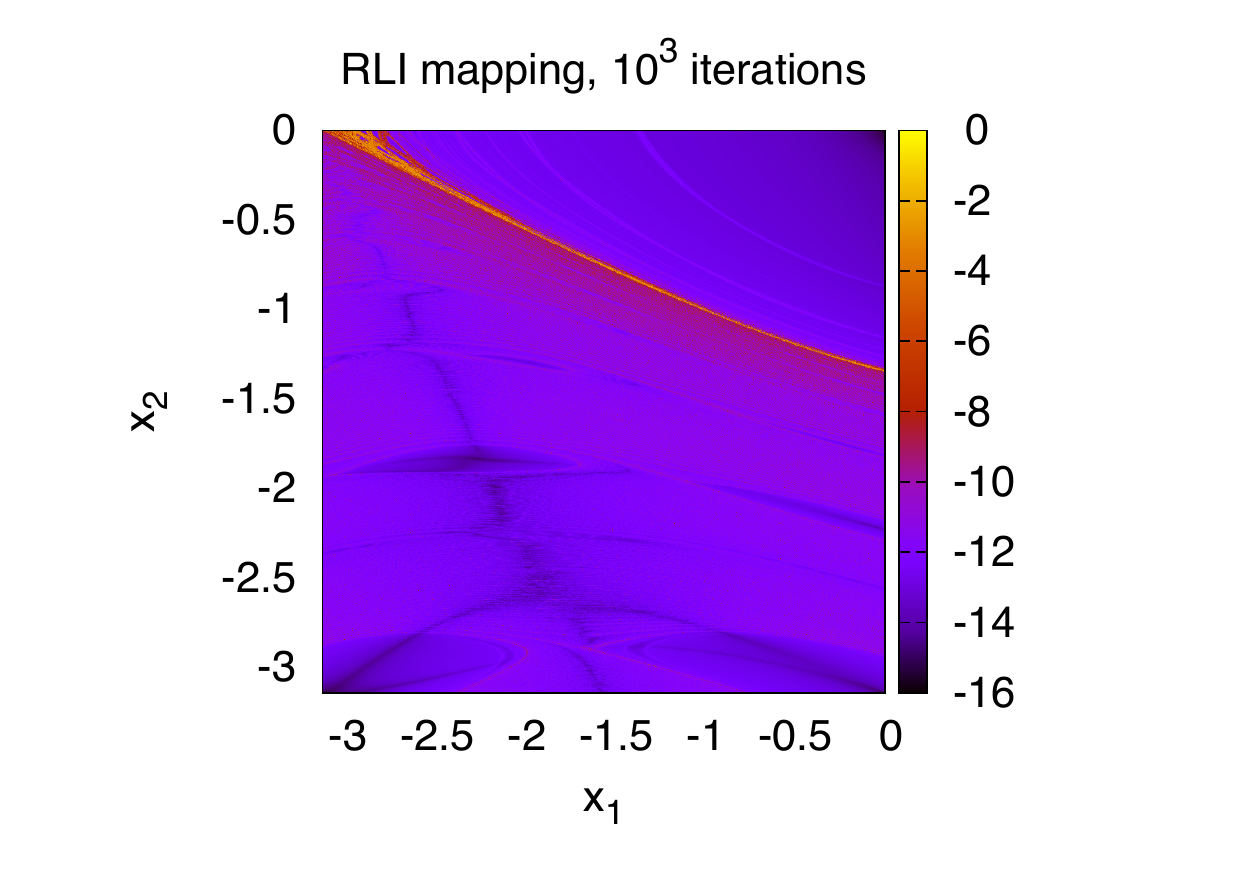}&
\hspace{-9mm}\includegraphics[scale=.40]{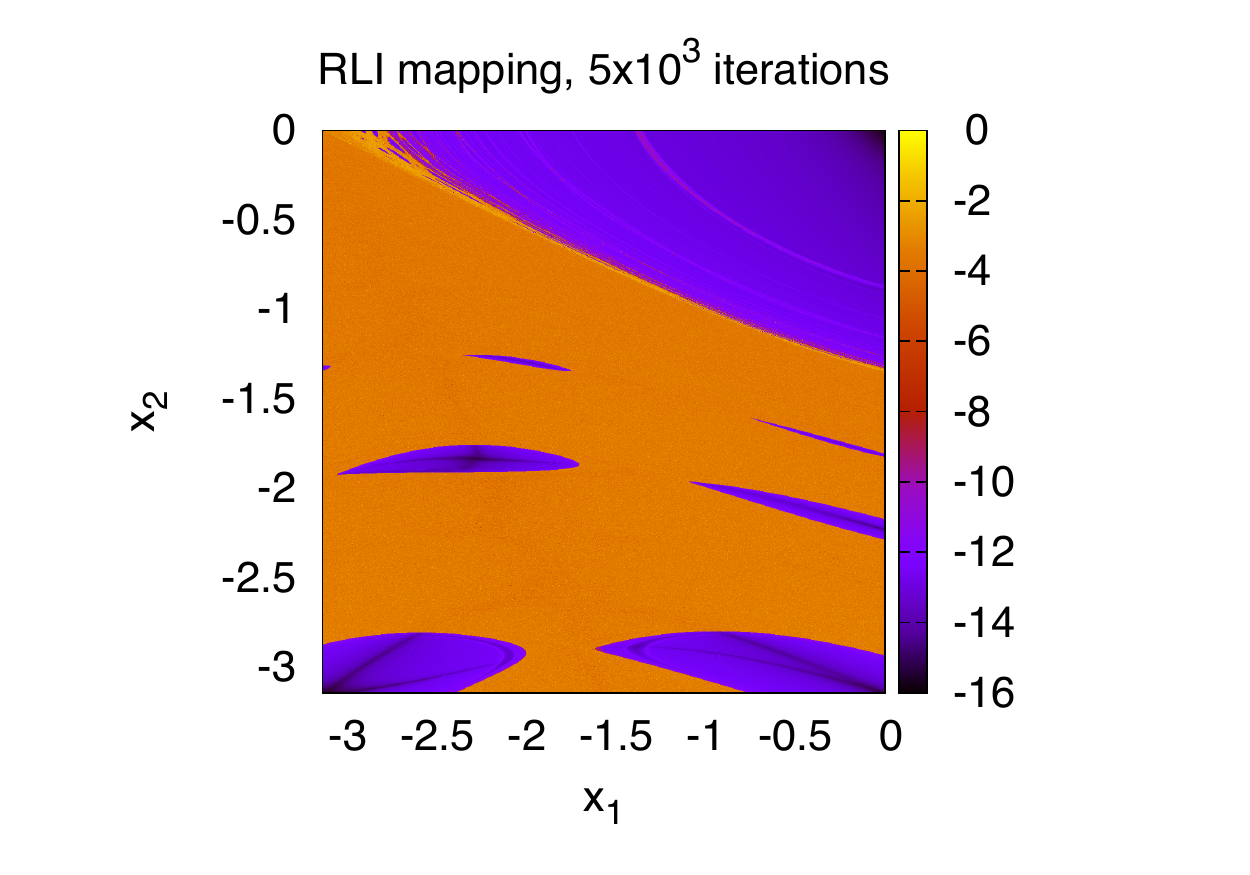}&
\hspace{-9mm}\includegraphics[scale=.40]{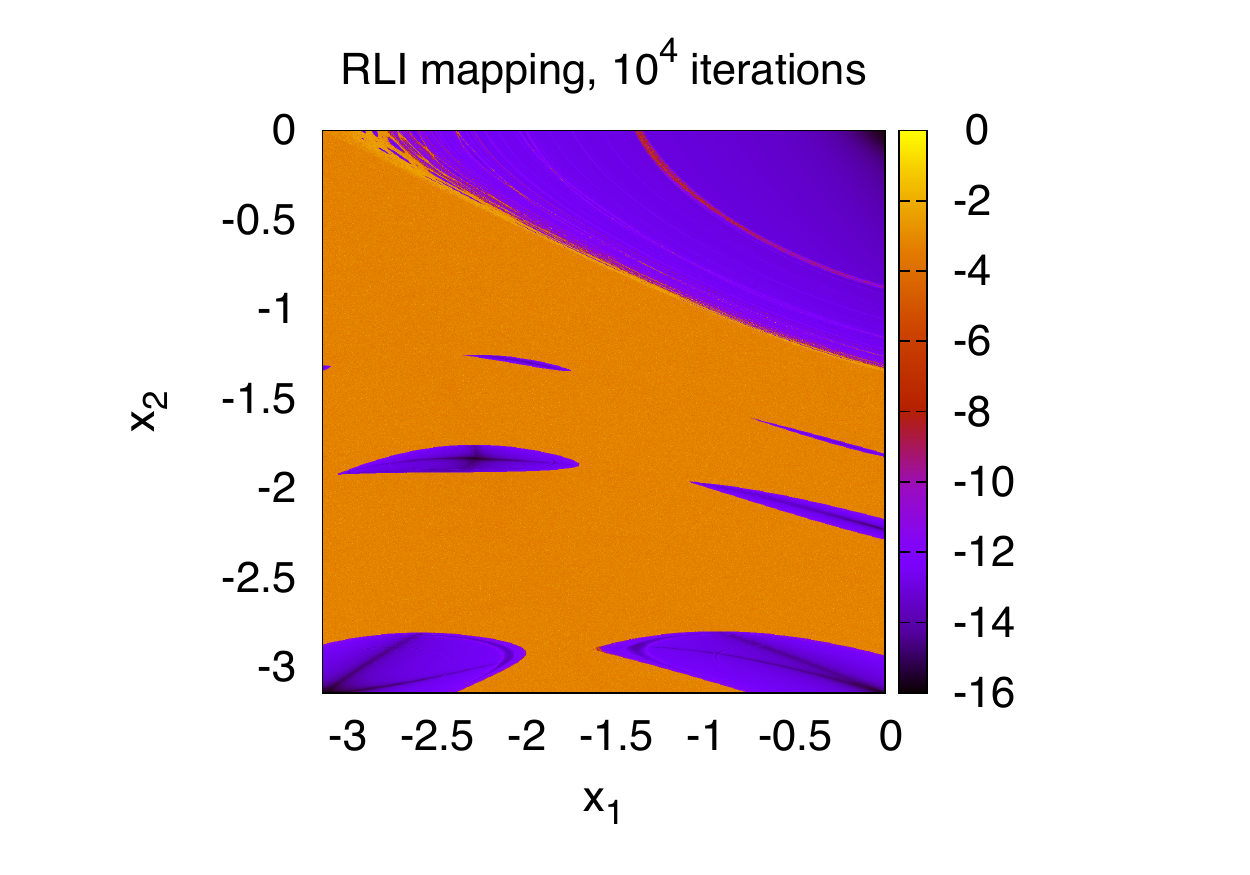}
\end{tabular}
\caption{The RLI mapping on gray--scale (color online) plots composed of $10^6$ initial conditions, for $10^3$ (left), $5\times10^3$ (middle) and $10^4$ (right) iterations.}
\label{time-rlimapping}
\end{figure}

The large number of iterations used in the experiments deserves a further explanation. In Fig. \ref{time-rlimapping}, we present the RLI mappings for $10^3$ (left panel), $5\times 10^3$ (middle panel) and $10^4$ (right panel) iterations. The RLI mapping corresponding to $10^3$ iterations presents a very noisy phase space portrait probably due to a combination of a poor election of the initial deviation vectors (see for instance \cite{B05,BBB09}) and the short number of iterations. It is also clear from the figure that the phase space portrait presents a stable picture after $5\times10^3$ iterations. Furthermore, increasing the number of iterations helps to resolve very sticky orbits but no further advantage is observed. Thus, we iterate the map $10^5$ times in order to distinguish the most sticky regions. 

\paragraph{Initial conditions inside high--order resonances}

\begin{figure}[t]
\sidecaption[t]
\begin{tabular}{cc}
\includegraphics[scale=.46]{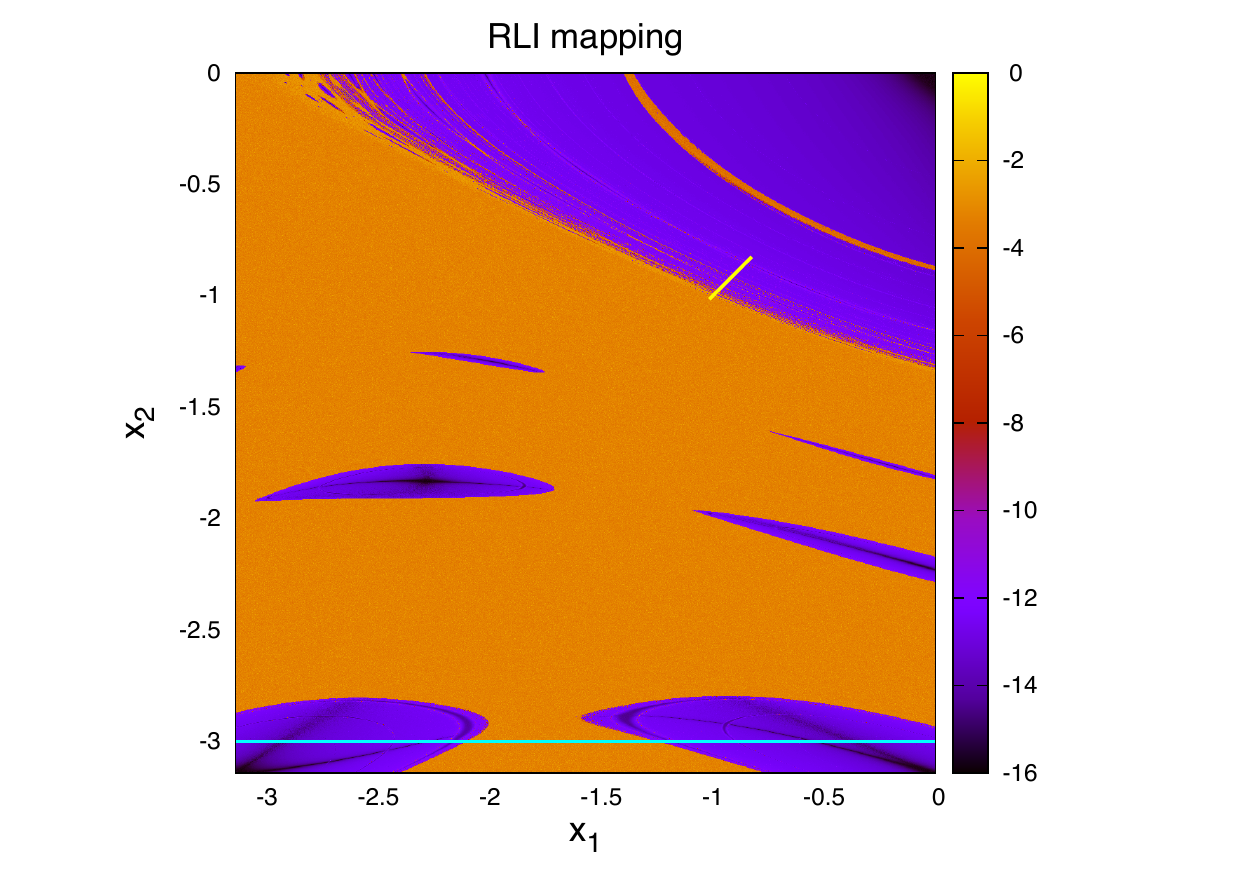}&
\includegraphics[scale=.46]{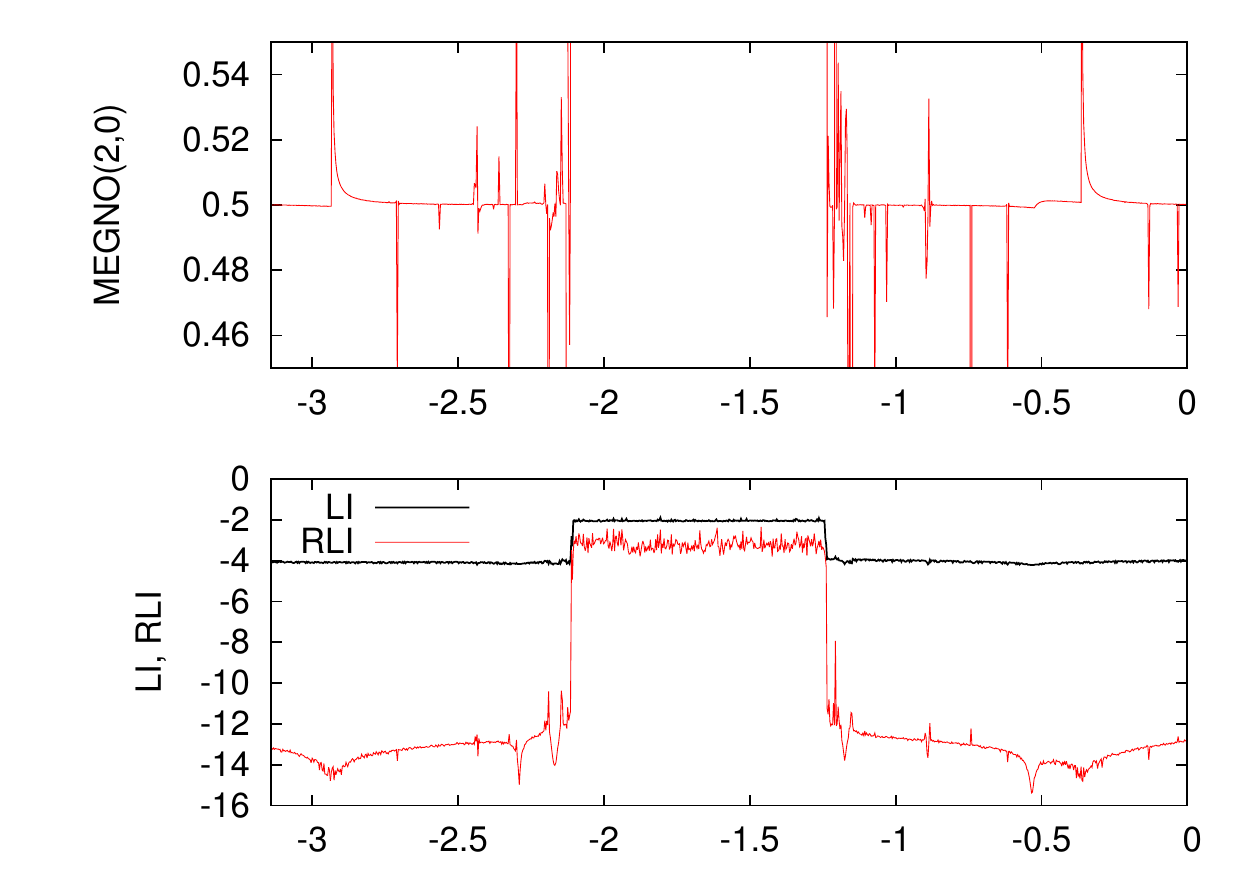}
\end{tabular}
\caption{(a) (left): The RLI mapping on gray--scale (color online) plot composed of $10^6$ initial conditions, for $10^5$ iterations (in logarithmic scale); (b) (right): The MEGNO(2,0), the LI and the RLI final values for $10^3$ initial conditions along the line $x_2=-3$, for $10^5$ iterations (the LI and the RLI final values are in logarithmic scale).}
\label{insideresonance}
\end{figure}

In Fig. \ref{insideresonance}(a) we present the RLI mapping for a region of the phase space corresponding to the 4D mapping and is composed of $10^6$ initial conditions. The main resonance as well as the high--order resonances are clearly depicted in dark gray (i.e. small values of the RLI) while the stochastic layers inside the main stability island and the chaotic sea are depicted in light gray (large values of the RLI). We show an horizontal line of initial conditions $(x_1\in[-\pi,0]$, $x_2=-3$, $x_3=0.5$ and $x_4=0)$ used in the following experiment to compare the performances of the RLI with the mostly used variational indicator, the LI, and with the MEGNO(2,0), which is faster than the LI and which is also a reliable indicator. With a diagonal segment we depict the initial conditions $(x_1=x_2\in[-1.03,-0.8]$, $x_3=0.5$ and $x_4=0)$ used in the experiment of next section. 

In Fig. \ref{insideresonance}(b), we compare the performances of the RLI, the LI and the MEGNO(2,0) on $10^3$ equidistant initial conditions lying on the horizontal line that crosses the high--order resonances in Fig. \ref{insideresonance}(a). This figure clearly shows that the RLI unzips the hidden structure inside the high--order resonances better than the LI. Furthermore, the RLI and the MEGNO(2,0) reveal similar structures (the Y-range of the MEGNO(2,0) has been centered on the threshold and shortened to amplify the details of the revealed structure). For further discussions on the experiment, refer to \cite{MDCG2011}.

On behalf of the previous experiments the RLI is not only more reliable than the LI to reveal small scale structures, but also an accurate indicator to describe a large array of initial conditions. 

\paragraph{Sticky orbits}

Sticky orbits are the most difficult type of orbit to characterize by a variational indicator. Thus, we further analyze the identification of this type of orbits (Sect. \ref{subseq:3}) to study the performance of the RLI. 

\begin{figure}[t]
\sidecaption[t]
\begin{tabular}{cc}
\includegraphics[scale=.46]{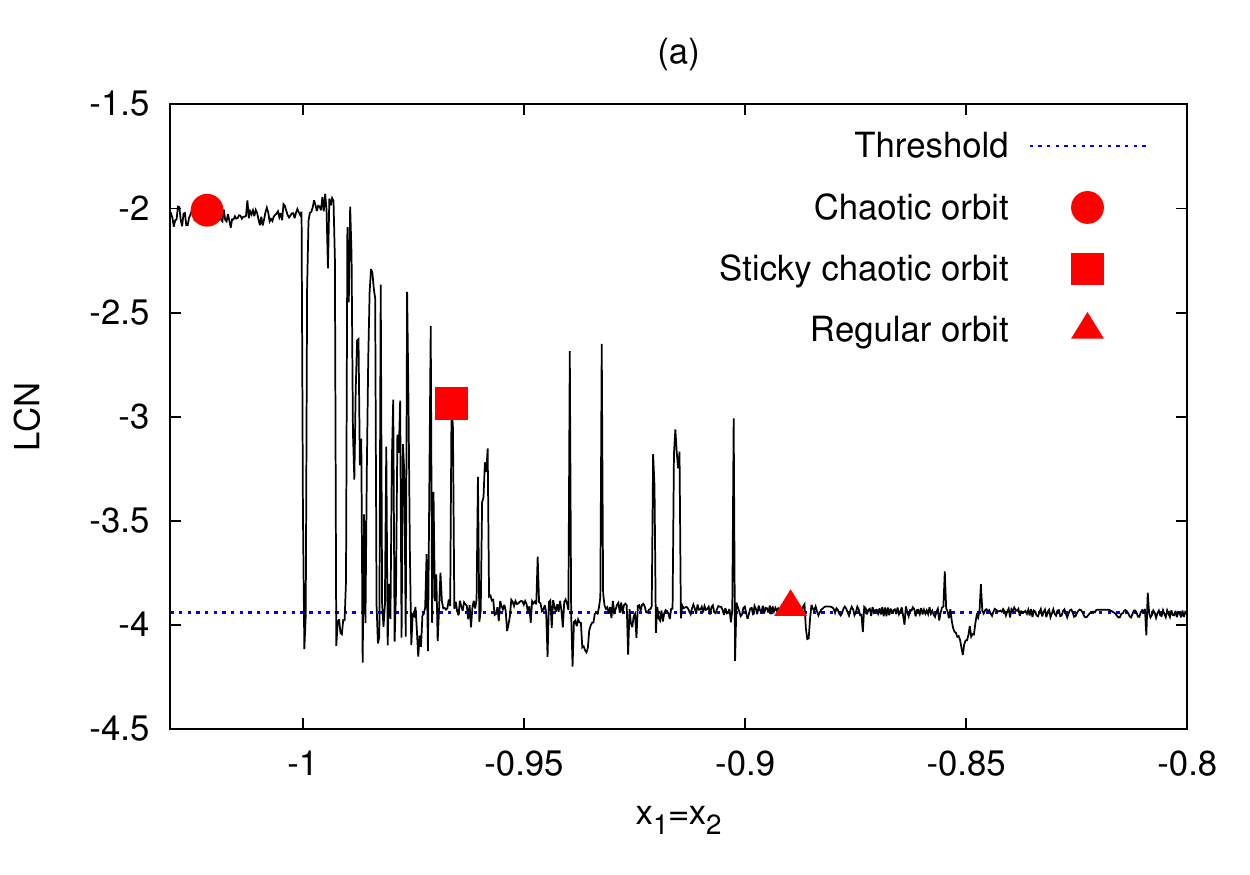}& 
\includegraphics[scale=.46]{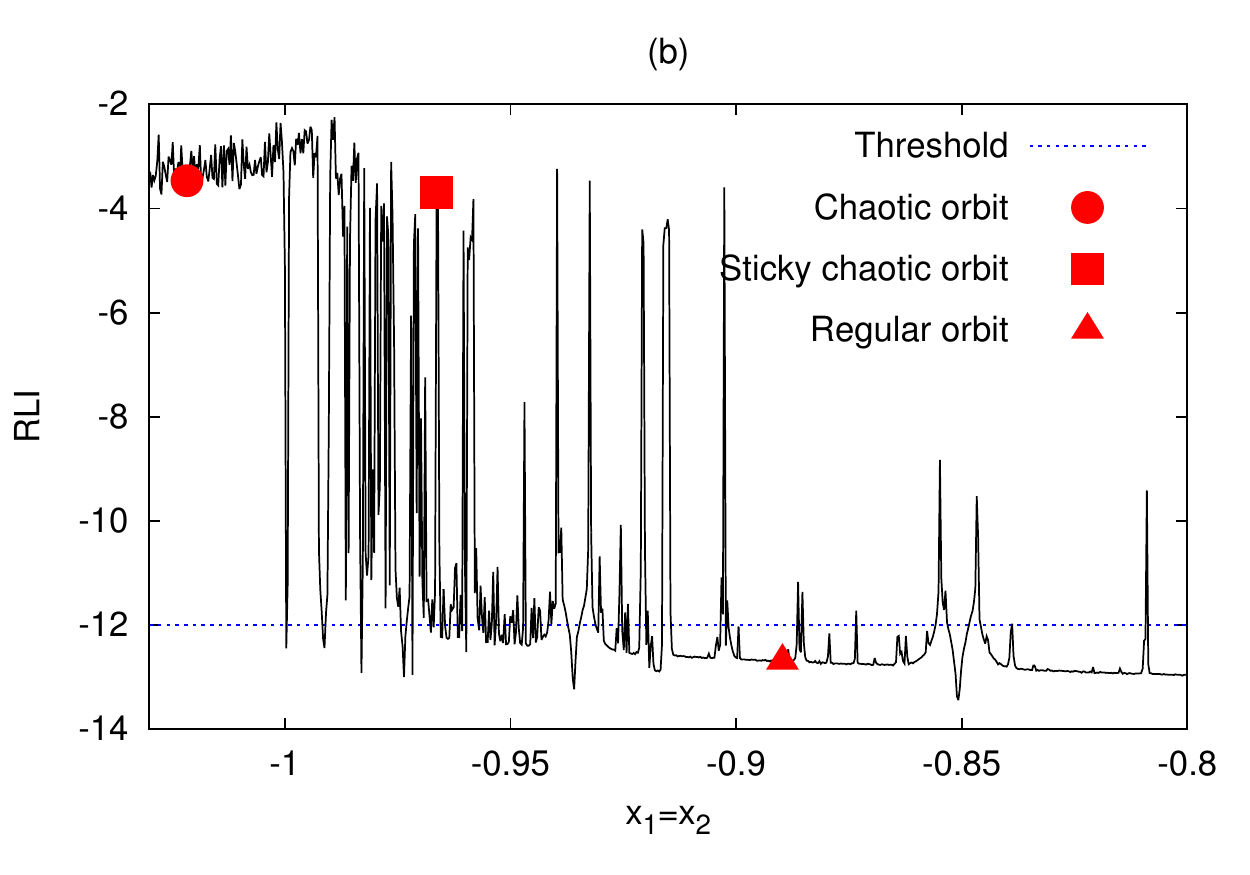}\\& 
\includegraphics[scale=.46]{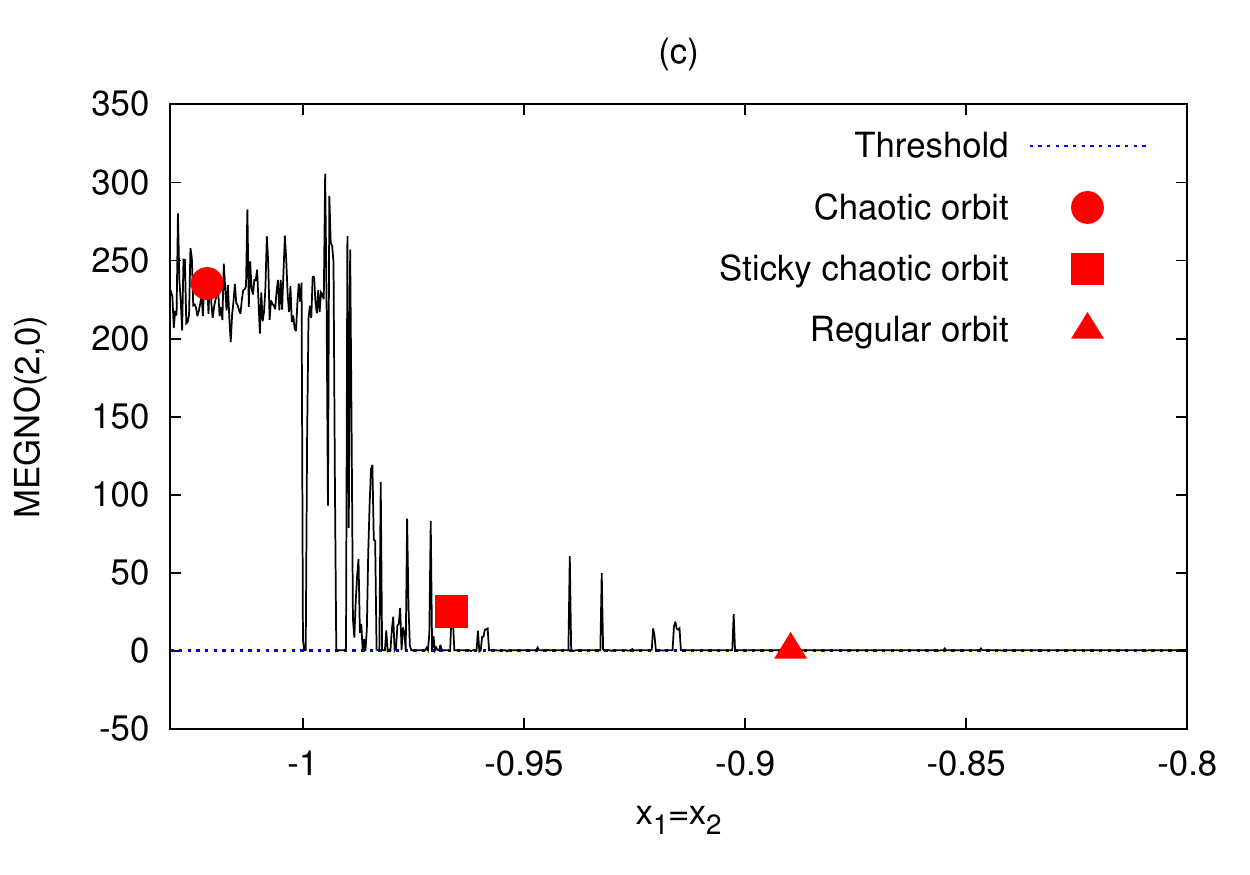}
\end{tabular}
\caption{Sticky region inside the interval (-1.03,-0.8) for $10^5$ iterations and for three indicators: (a) the LI, (b) the RLI and (c) the MEGNO(2,0). The representative orbits are depicted with points of different types. The thresholds are depicted with a dashed line (the LI and the RLI final values are in logarithmic scale). See text for details. The figures were taken from \cite{MDCG2011} and slightly modified.}
\label{stickyamp}
\end{figure}

In Fig. \ref{stickyamp}, we show the sticky region enclosed in the interval (-1.03,-0.8) (and depicted earlier in Fig. \ref{insideresonance}(a) with a diagonal segment) in terms of the final values of the same indicators previously used. We also point out the final values of three representative orbits, two chaotic orbits (one of them which is sticky chaotic) and a regular orbit. The thresholds are also included.

In Fig. \ref{stickyamp}(a), the LI perfectly indentifies the three representative orbits and their domains. In Fig. \ref{stickyamp}(b), some sticky and chaotic orbits share the same RLI final values ($\sim10^{-3.5}$) which hide the different levels of chaoticity (such is the case of the representative chaotic and sticky chaotic orbits, both of them have similar RLI values). This is not the case for the other indicator shown in the figure: the MEGNO(2,0) on panel (c). The MEGNO(2,0) has completely different final values for the sticky and the chaotic orbits.

The RLI shows a reliable performance revealing the global characteristics of the region, such as the regular domain (where we find the representative regular orbit) and some small high--order resonances (e.g. $x_1=x_2\sim -0.85$). Nevertheless, it does not distinguish the sticky from the chaotic orbit in the experiment (see \cite{MDCG2011} for further details). 

In the following section, we follow with the experiments in a scenario of completely different nature and much more complex than the HH model or the 4D mapping.

\subsubsection{The 3D NFW model}
\label{sec:3.3.2}

According to the current paradigm of a hierarchically clustering universe, large galaxies formed through the accretion and mergers of smaller objects. The imprints of such events should be well preserved in the outer stellar haloes where dynamical mixing processes are not significant in relatively short times (for instance, many stellar streams have been identified in the outer regions of the Milky Way \cite{IILS2001,MSWO2003,B2007}). Furthermore, this galaxy formation paradigm predicts that the centres of the accreted component of stellar haloes should contain the oldest products of accretion events \cite{C2010} in the formation history of the galaxy (such as the Milky Way), and therefore, this substructure might be hidden in its inner regions (e.g. close to the Solar neighbourhood). However, in order to study the  phase space portraits of stars in small volumes in the inner regions of the stellar haloes to quantify and classify the substructure, we need a model of the Dark Matter (DM) halo that hosts the galaxy.

In \cite{NFW1996,NFW1997} the authors introduced a universal density profile for DM haloes (i.e. haloes with masses ranging from dwarf galaxy haloes to those of major clusters): the NFW profile. However, in Cold Dark Matter (CDM) cosmologies DM haloes are not spherical. Furthermore, numerical simulations suggest that their shape vary with radius. In \cite{VWHS2008} the authors have built a triaxial extension of the NFW profile, resembling a triaxial DM halo; the corresponding potential is the so--called NFW model. Therefore, the NFW model is a triaxial potential used in galactic dynamics associated with equilibrium density profiles of DM haloes in CDM cosmologies.

In a forthcoming paper we study the phase space portraits of stellar particles inside solar neighbourhood--like volumes to gain insights about the formation history of Milky Way--like galaxies. Hereinafter, we use some results from that investigation to demonstrate the reliability of the RLI on a rather complex model.

The solar neighbourhood volume is a sphere of 2.5 Kpc of radius located at 8 Kpc from the center of the NFW model (denoted by $\Phi_{\rm N}$ in the following equation):

\begin{equation*}
\Phi_{\rm N}=-\frac{A}{r_p}\ln \left( 1+\frac{r_p}{r_s}\right)\ ,
\end{equation*}
where $A$ is the constant: 

\begin{equation*}
A=\frac{G\,M_{200}}{\ln\left(1+C_{NFW}\right)-C_{NFW}/\left(1+C_{NFW}\right)}\ ,
\end{equation*}

with $G$, the gravitational constant, $M_{200}$, the virial mass of the DM halo and $C_{NFW}$, the concentration parameter used to describe the shape of the density profile. Besides, $r_p$ follows the relation:

\begin{equation*}
r_p=\frac{(r_s+r)r_e}{r_s+r_e}\ ,
\end{equation*}
where $r_s$ is the scale radius defined by dividing the virial radius of the DM halo by $C_{NFW}$. The scale radius represents a transition scale between an ellipsoidal and a near spherical shape of the $\Phi_{\rm N}$. The $r_e$ is an ellipsoidal radius:

\begin{equation*}
r_e=\sqrt{\left(\frac{x}{a}\right)^2 + \left(\frac{y}{b}\right)^2 +
\left(\frac{z}{c}\right)^2}\ ,
\end{equation*}
where $b/a$ and $c/a$ are the principal axial ratios with $a$ the major axis and where we require $a^2+b^2+c^2=3$ (see \cite{VWHS2008}). The values of the constants used in the following experiment are taken at redshift $z=0$ and listed in Table \ref{const} (see \cite{GHCFNW2013} and references therein for further details on the model). 

\begin{table}\centering
\caption{Constants used for the $\Phi_{\rm N}$ potential}
\label{const}
\begin{tabular}{p{1cm}p{3cm}}
\hline\noalign{\smallskip}
 $A$ & 4158670.1856267899 \\
 $r_s$ & 19.044494521343964 \\
 $a$ & 1.3258820840000000 \\
 $b$ & 0.86264540200000000 \\
 $c$ & 0.70560584600000000 \\
\noalign{\smallskip}\hline\noalign{\smallskip}
\end{tabular}
\end{table}

In order to begin the study of the (6 dimensional) regions of interest in the phase space of the NFW model, we needed to restrain some of the variables  that defined the original sample of $22500$ initial conditions. We fixed the positions of the particles to the centre of the solar neighbourhood. Then, the stellar particles had the following positions at the begining of the simulation: $x=8$, $y=0$ and $z=0$ (in [Kpc]). The initial velocity in the polar axis ($v_z$) was restrained to the value $-250$ in [km s$^{-1}$]. The energy ($E$) was restrained to the interval $(E_{mb},E_{lb})$ with $E_{mb}\sim-195433$ the energy of the most bound particle, and $E_{lb}\sim-59293$ (in [M$_\odot$ Kpc$^2$ Gyr$^{-2}$] with [M$_\odot$] the mass in solar mass units) the energy of the least bound particle of the sample. The angular momentum ($L_z$) was restrained to the interval $(-2000,2000)$ in [Kpc km s$^{-1}$]. We integrated the initial conditions for different integration times (1 u.t. corresponds to $\sim$1 Gyr). 

\begin{figure}[t]
\sidecaption[t]
\begin{tabular}{cc}
\hspace{-18mm}\includegraphics[scale=.65]{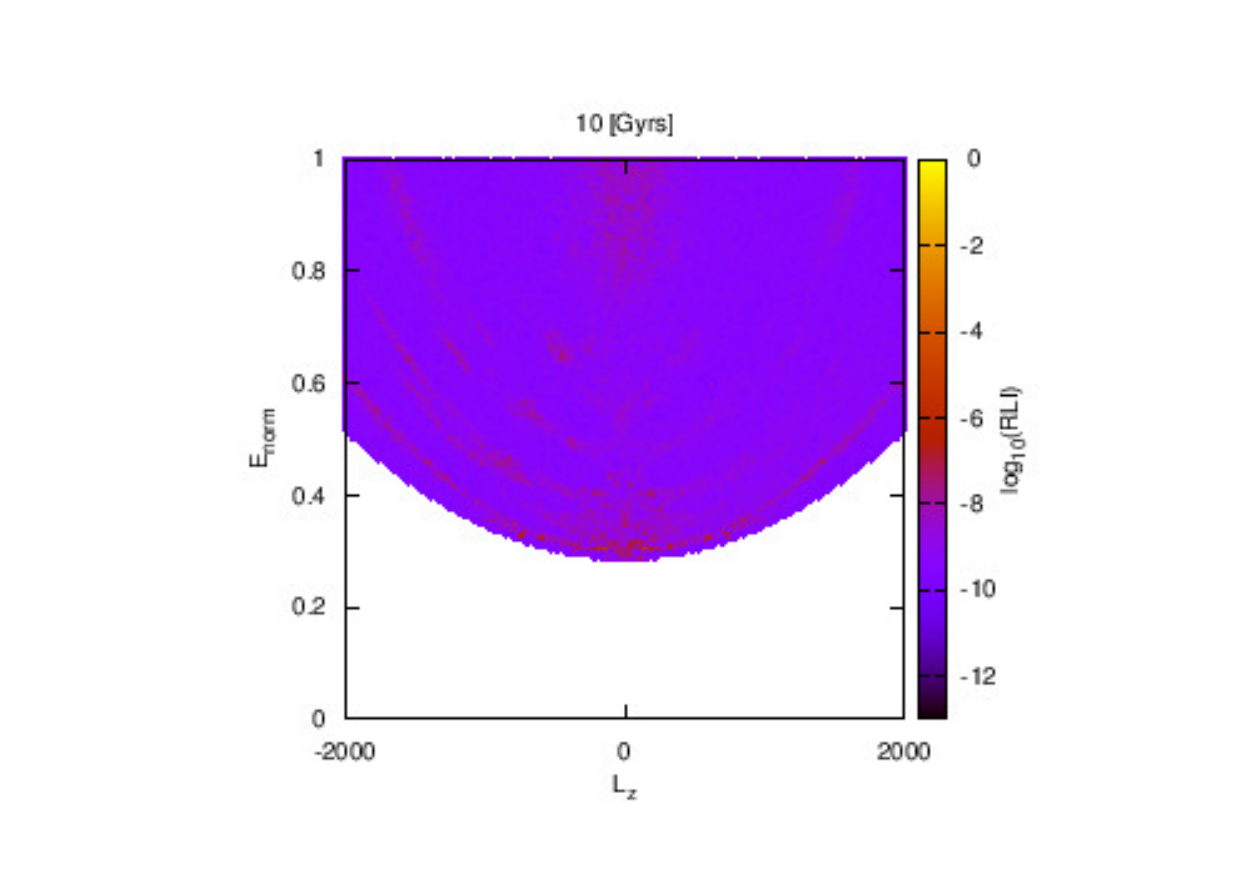}&
\hspace{-19mm}\includegraphics[scale=.65]{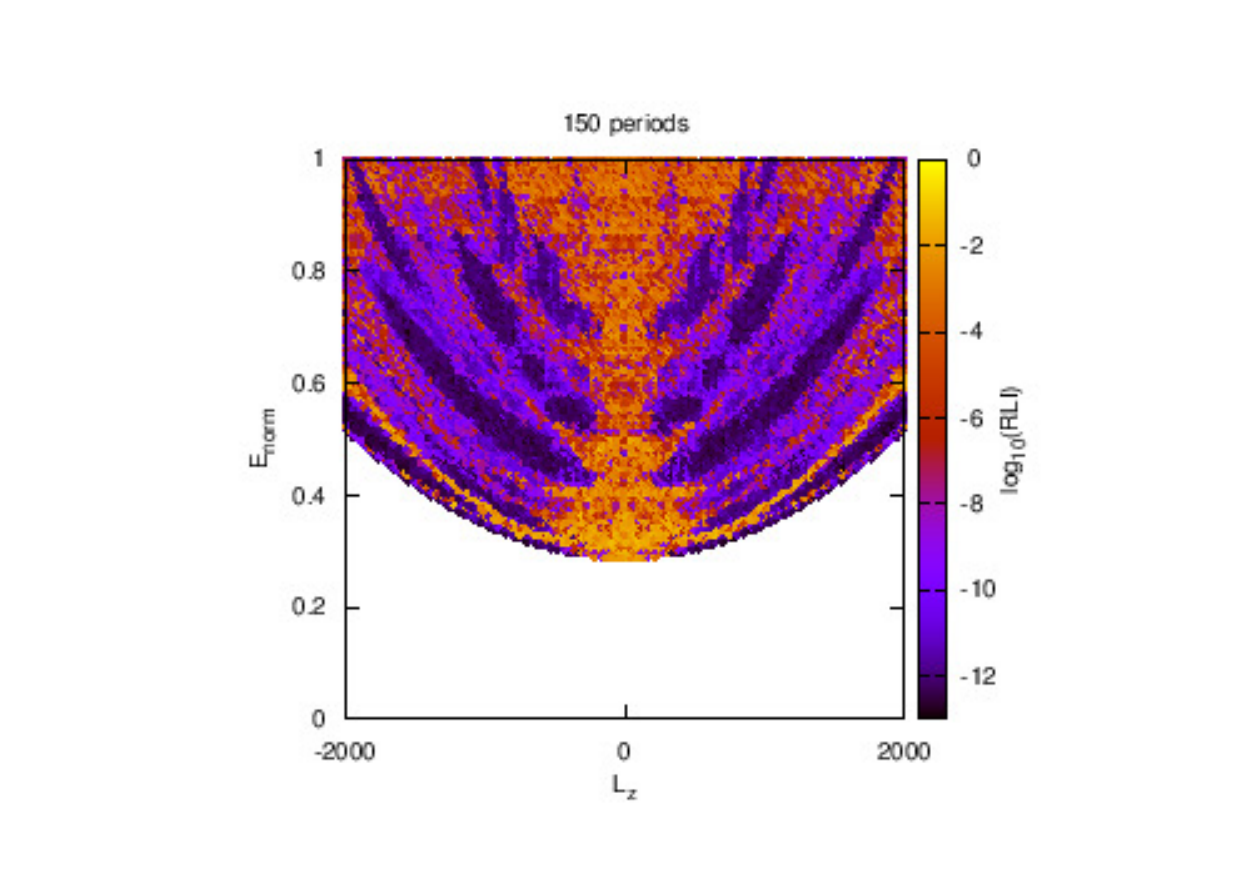}
\end{tabular}
\caption{Gray--scale (color online) plots of the RLI mapping for the velocity surface $v_z=-250$. (left): For a fixed integration time of 10 Gyrs. (right): For a fixed number (150) of radial periods. The values of the indicator are in logarithmic scale.}
\label{vz-250-pre}
\end{figure}

In Fig. \ref{vz-250-pre}, we present the phase space portraits given by the RLI for two different choices of the integration times. On the left panel, we integrate the orbits for a fixed time interval of the order of the Hubble time, i.e. 10 Gyrs. On the right panel, we integrate the orbits for a fixed number of radial periods. The radial period of the stellar orbits in galactic potentials such as the NFW model scales as $\sim E^{-3/2}$. Then, we integrate the orbits for 150 radial periods in order to have a stable portrait of the phase space.  It is evident that 10 Gyrs (left panel) is not enough to classify properly the orbits with the RLI (or any other indicator). Then, on the right panel the time interval used was $[\sim57,750]$ Gyrs where 750 Gyrs is enough time to set reliable values of the RLI for the least bound particle of the sample. However, the most sticky regions are not clearly depicted yet. Therefore, in the following experiment we choose to scale the integration time linearly with the energy of the orbit. The linear relation between the computed integration times and the energy overestimates the former for the most bound particles. Indeed, the time interval used for the experiment was $[\sim204,750]$ Gyrs where the integration times are clearly larger than those applied with the $\sim E^{-3/2}$ scale. The larger integration times given by the linear scale improve the identification of the most sticky regions which helps to evaluate the performance of the indicators. 

\begin{figure}[t]
\sidecaption[t]
\begin{tabular}{cc}
\hspace{-18mm}\includegraphics[scale=.65]{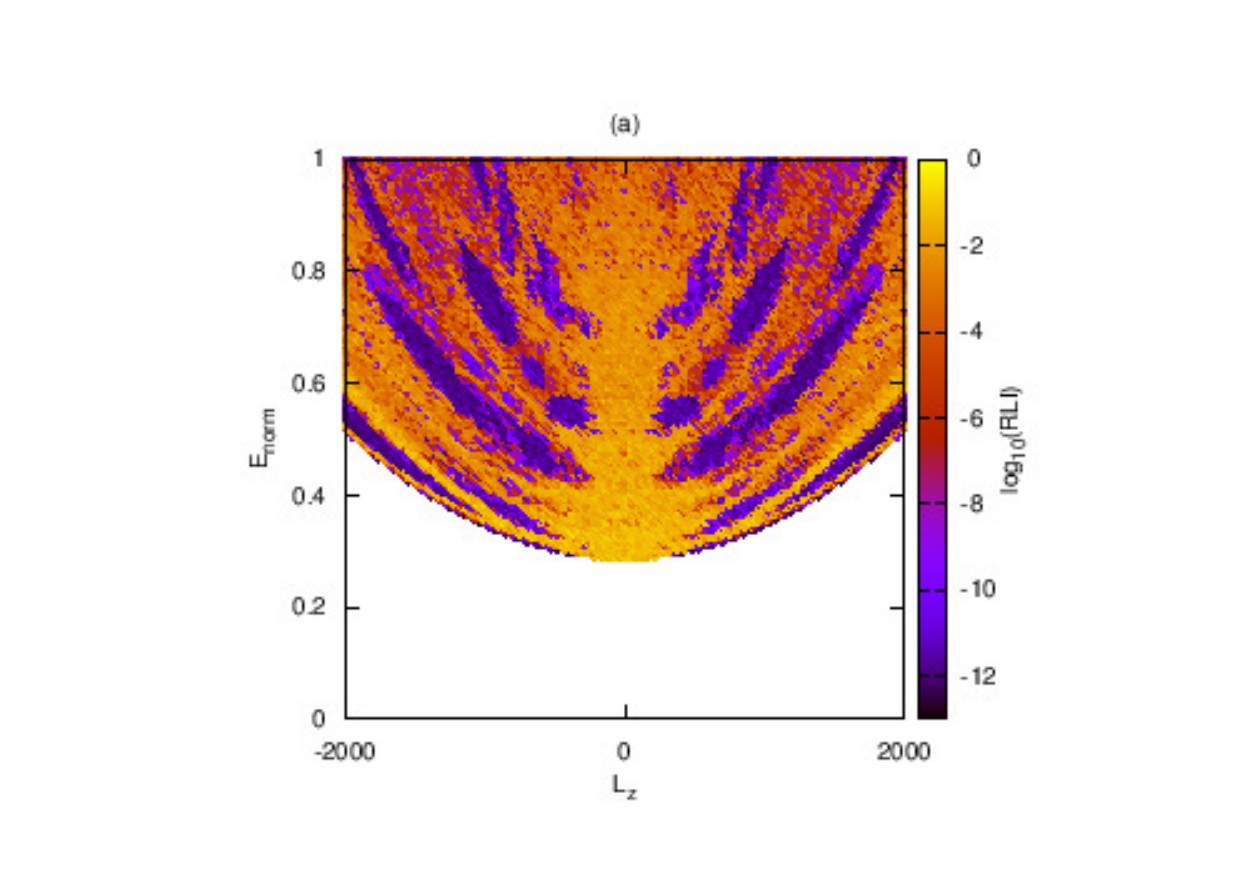}&\\
\hspace{-18mm}\includegraphics[scale=.65]{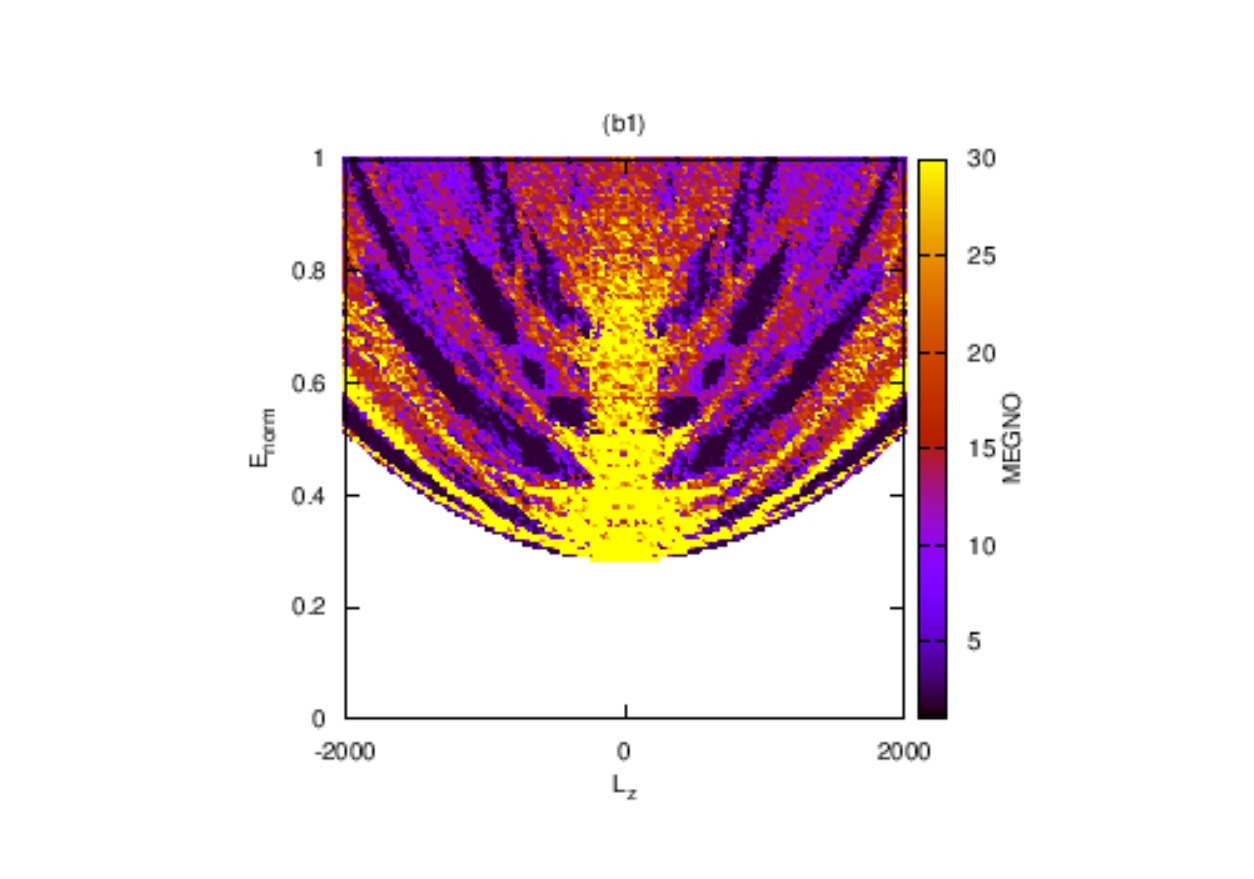}&
\hspace{-19mm}\includegraphics[scale=.65]{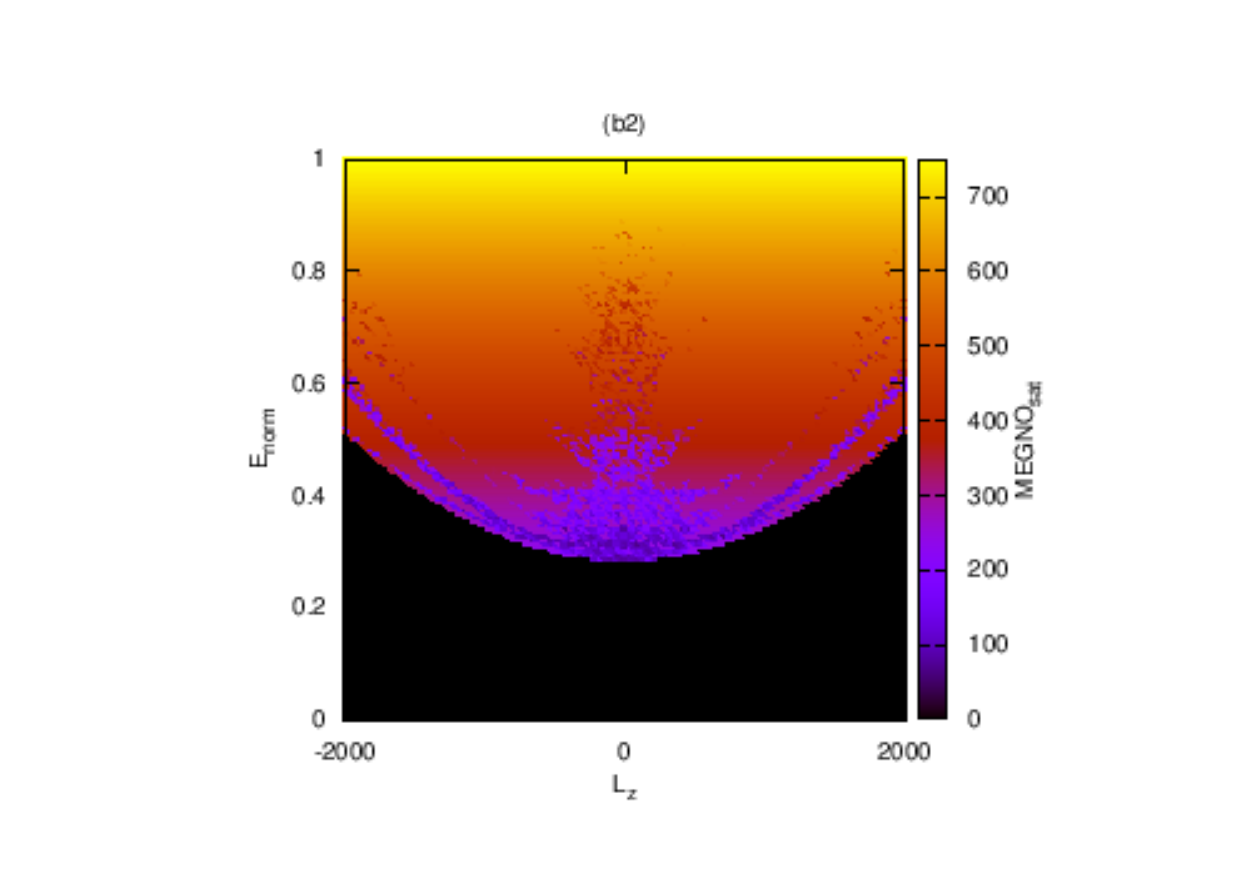}
\end{tabular}
\caption{Gray--scale (color online) plots for the velocity surface $v_z=-250$. (a) The RLI mapping (the values of the indicator are in logarithmic scale), (b1) the MEGNO mapping and (b2) the MEGNO$_{sat}$ mapping.}
\label{vz-250a}
\end{figure}

\begin{figure}[t]
\sidecaption[t]
\begin{tabular}{cc}
\hspace{-15mm}\hspace{-5mm}\includegraphics[scale=.65]{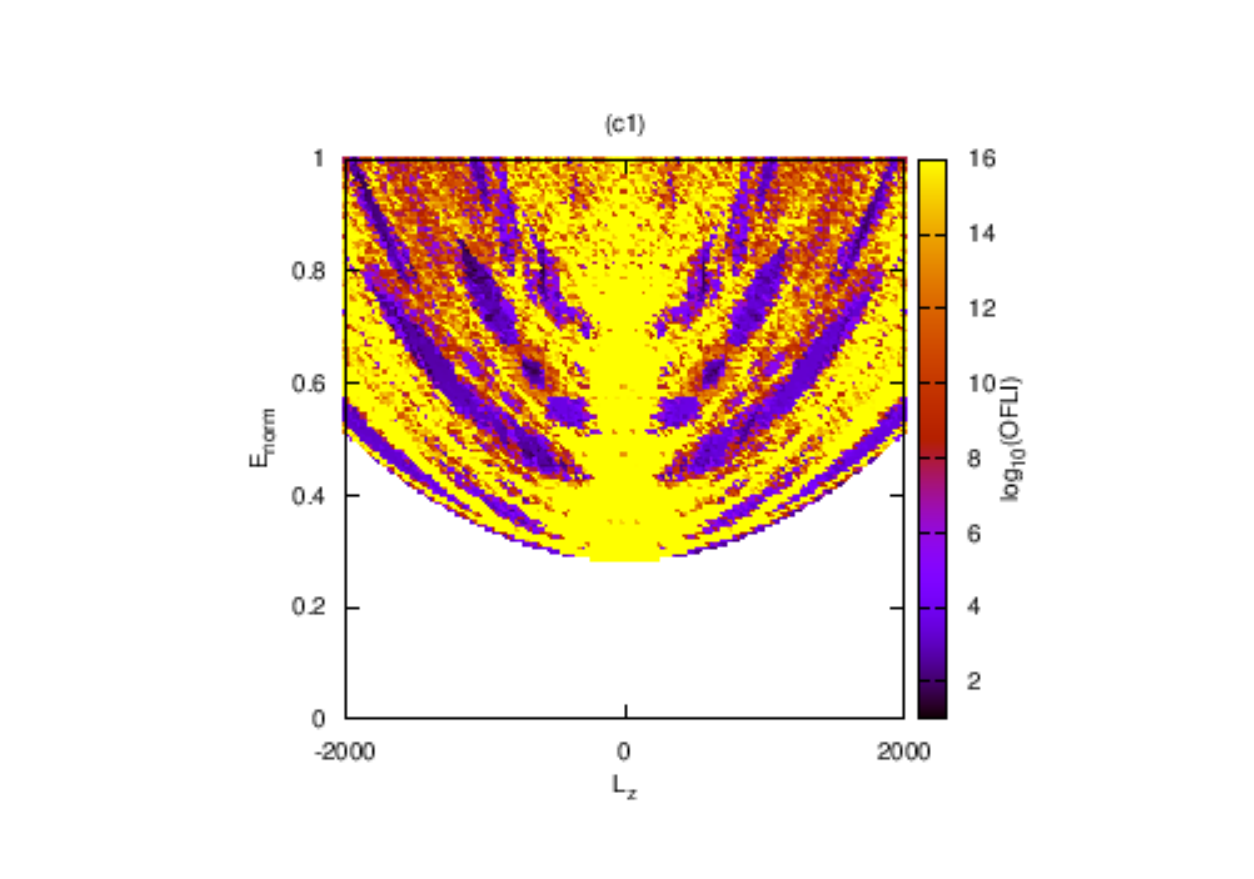}&
\hspace{-10mm}\hspace{-5mm}\includegraphics[scale=.65]{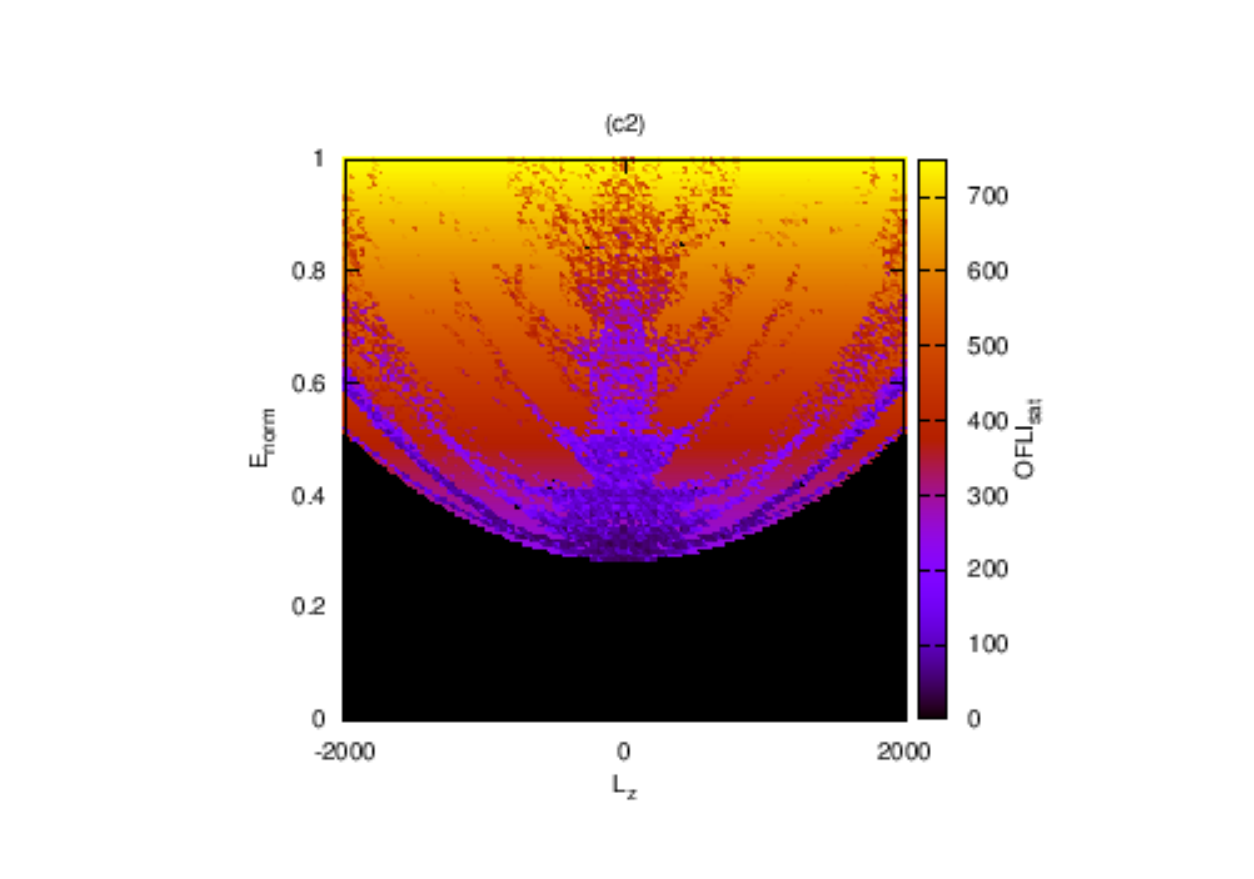}\\
\hspace{-15mm}\hspace{-5mm}\includegraphics[scale=.65]{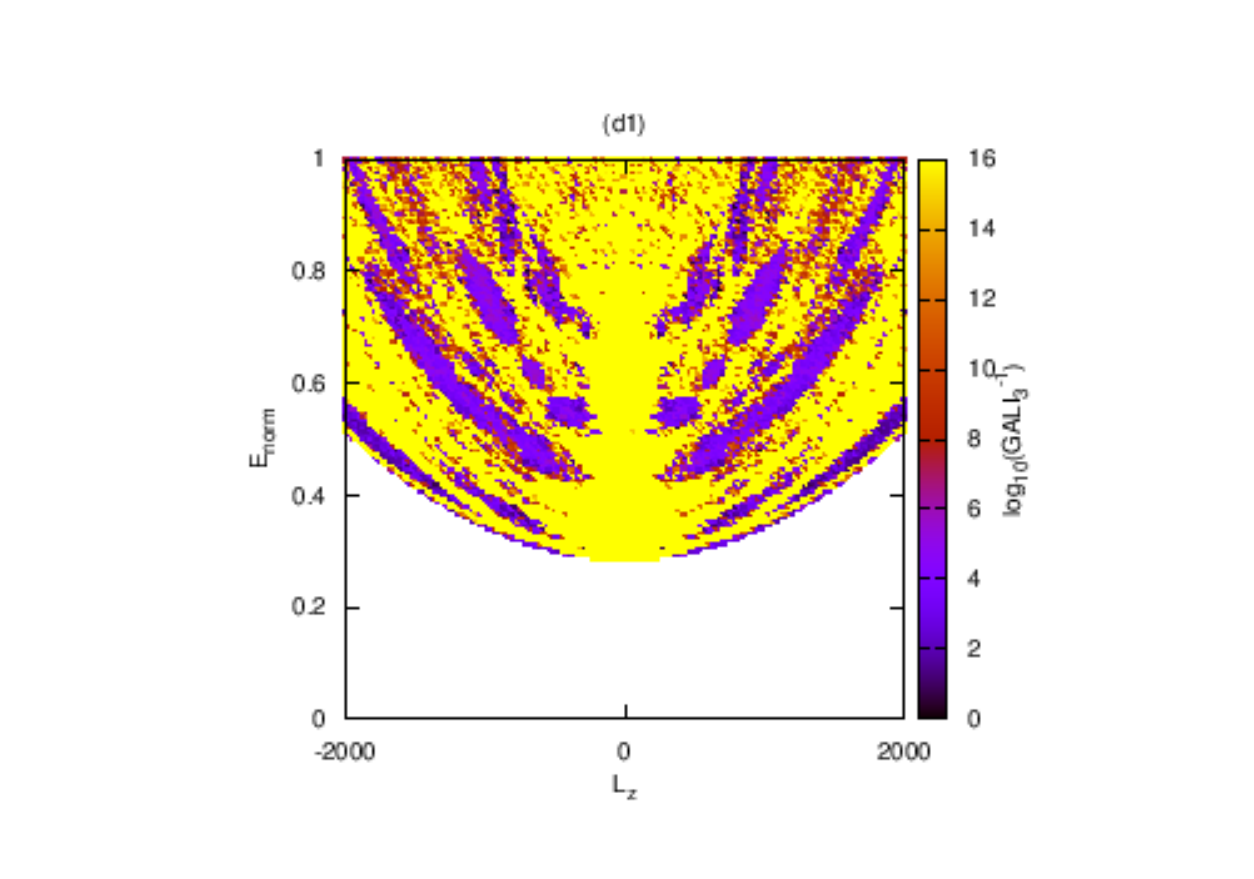}&
\hspace{-10mm}\hspace{-5mm}\includegraphics[scale=.65]{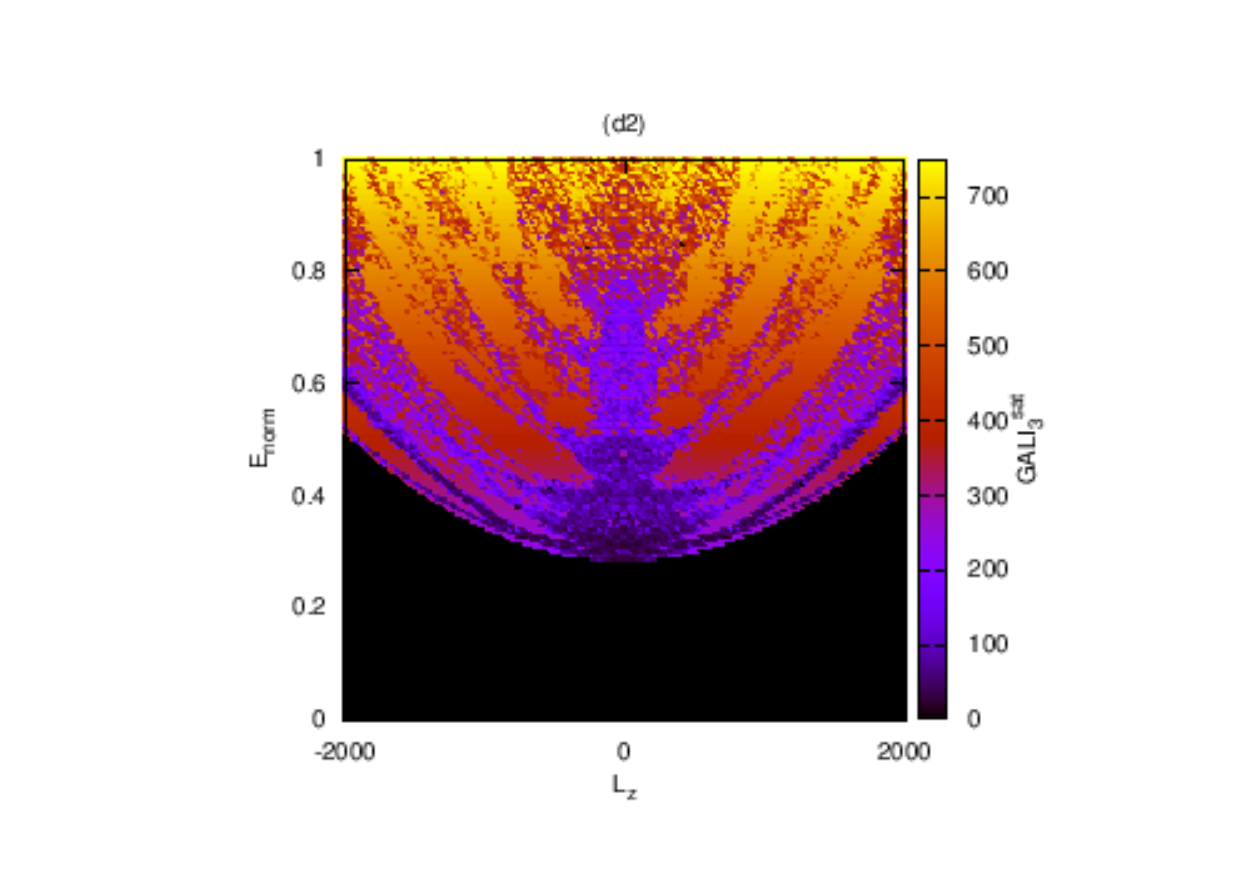}
\end{tabular}
\caption{Gray--scale (color online) plots for the velocity surface $v_z=-250$. (c1) The OFLI mapping and (c2) the OFLI$_{sat}$ mapping, (d1) the GALI$_3$ mapping and (d2) the GALI$_3^{sat}$ mapping (the final values of the indicators are in logarithmic scale).}
\label{vz-250b}
\end{figure}

On the left panels of Figs. \ref{vz-250a} and \ref{vz-250b} we present the gray--scale plots of the final values of four different chaos indicators in the $(E_{norm}, L_z)$ plane\footnote{The $E_{norm}$ is the normalized energy: $\left(E-E_{mb}\right)/\left(E_{lb}-E_{mb}\right)$.}: the RLI and the MEGNO; panels (a) and (b1) of Fig. \ref{vz-250a}, respectively; the OFLI and the 1/GALI$_3$: panels (c1) and (d1) of Fig. \ref{vz-250b}, respectively. The GALI$_3$ is inverted in the color--scale plots to make the comparison of the portraits with the other indicators easier.

In many situations it is useful to define a saturation value by which the chaos indicator ``saturates''. For instance, the OFLI and the GALI$_3$ for chaotic motion increases or decreases exponentially fast, respectively. Then, if the chaotic nature of an orbit is well characterized by the OFLI when the indicator reaches $10^{16}$ or by the GALI$_3$ when it reaches the computer's precision ($10^{-16}$), the computation should be stopped. Hence, the values $10^{16}$ and $10^{-16}$ can be used as saturation values for the OFLI and the GALI$_3$, respectively. Another example is the MEGNO: the MEGNO has an asymptotic value for regular orbits, 2 (see Table \ref{tablethreshold}), and increases linearly for chaotic orbits. Then, if the MEGNO reaches the value 30, the orbit is undoubtfully chaotic and it is worthless to continue the computation of the indicator. Then, the value 30 can be used as a saturation value for the MEGNO. The time of saturation, that is the time by which the indicator saturates, it is a quantity useful in recovering the chaoticity levels in the chaotic domain: the smaller the value of the time of saturation, the more chaotic the orbit (see \cite{SBA2007,MDCG2011}). Finally, if the indicator saturates (i.e. the indicator reaches its saturation value), the integration times will be the times of saturation, but, if the indicator does not saturate, the integration times will be the final integration times.

On the right panels of Figs. \ref{vz-250a} and \ref{vz-250b} we present gray--scale plots of the integration times used for three of the four indicators above mentioned. On panel (b2) in Fig. \ref{vz-250a} we present the integration times for the MEGNO or MEGNO$_{sat}$ and in Fig. \ref{vz-250b}: panels (c2) and (d2), the integration times for the OFLI and the GALI$_3$, or OFLI$_{sat}$ and GALI$_3^{sat}$, respectively.

The discussion below is not intended to analyze the dynamics of the system, which is the aim of a future work. Our goal here is to demonstrate that the performance of the RLI in this complex system is as good as the performances of the other three wide--spread indicators. 

On the left panels of Figs. \ref{vz-250a} and \ref{vz-250b}, we can clearly see the great level of coincidence among the phase space portraits of the four indicators. The regular component is composed of symmetrical structures around the $L_z$ axis and the four indicators represent these structures with very similar shapes, sizes and shades of gray. This shows that the indicators do not only agree in the location, extension and shape of the domains of regular motion, but also in the description of these domains.    

The chaotic domain is described equivalently by the four indicators and their corresponding times of saturation. For instance, the RLI (Fig. \ref{vz-250a}(a)) shows that the chaoticity (light gray) is inversely proportional to $E_{norm}$ and does not depend on the $L_z$. That is, more bound orbits are more chaotic orbits. We arrive at the same (trivial) conclusion with the information given by the integration times of the other three indicators (Figs. \ref{vz-250a}(b2), \ref{vz-250b}(c2) and \ref{vz-250b}(d2)). The MEGNO (Fig. \ref{vz-250a} (b1)) shows an uniform and almost white color for the chaotic domain. This implies that the saturation value (30) has been reached by these chaotic orbits and thus, no further structure is revealed. However, the MEGNO$_{sat}$ (Fig. \ref{vz-250a}(b2)) shows such structure (surrounded by orbits that did not saturate)\footnote{Remember that the integration time varies with the $E_{norm}$, which explains the transition from dark to light gray in the background of the plots on the left side of Figs. \ref{vz-250a} and \ref{vz-250b}. Also the time of integration is fixed to 0 where there are not initial conditions.}. This structure revealed by the MEGNO$_{sat}$ shows that the times of saturation are directly proportional to $E_{norm}$ or, once again, that chaoticity is inversely proportional to $E_{norm}$ and also does not depend on the $L_z$. Similar conclusions can be drawn from Fig. \ref{vz-250b} for the OFLI (panel c1) and the OFLI$_{sat}$ (panel c2) and the GALI$_3$ (panel d1) and GALI$_3^{sat}$ (panel d2). However, the region composed of orbits that have reached the associated saturation values ($10^{16}$ and $10^{-16}$ for the OFLI and the GALI$_3$, respectively) within the interval of integration is now much extended. On the one hand, this region in Figs. \ref{vz-250b}(c1) and (d1) is depicted in an uniform and almost white color and thus, the structure cannot be revealed. On the other hand, the times of saturation in Figs. \ref{vz-250b}(c2) and (d2) fulfill the missing information.      

\begin{figure}[t]
\sidecaption[t]
\includegraphics[scale=.46]{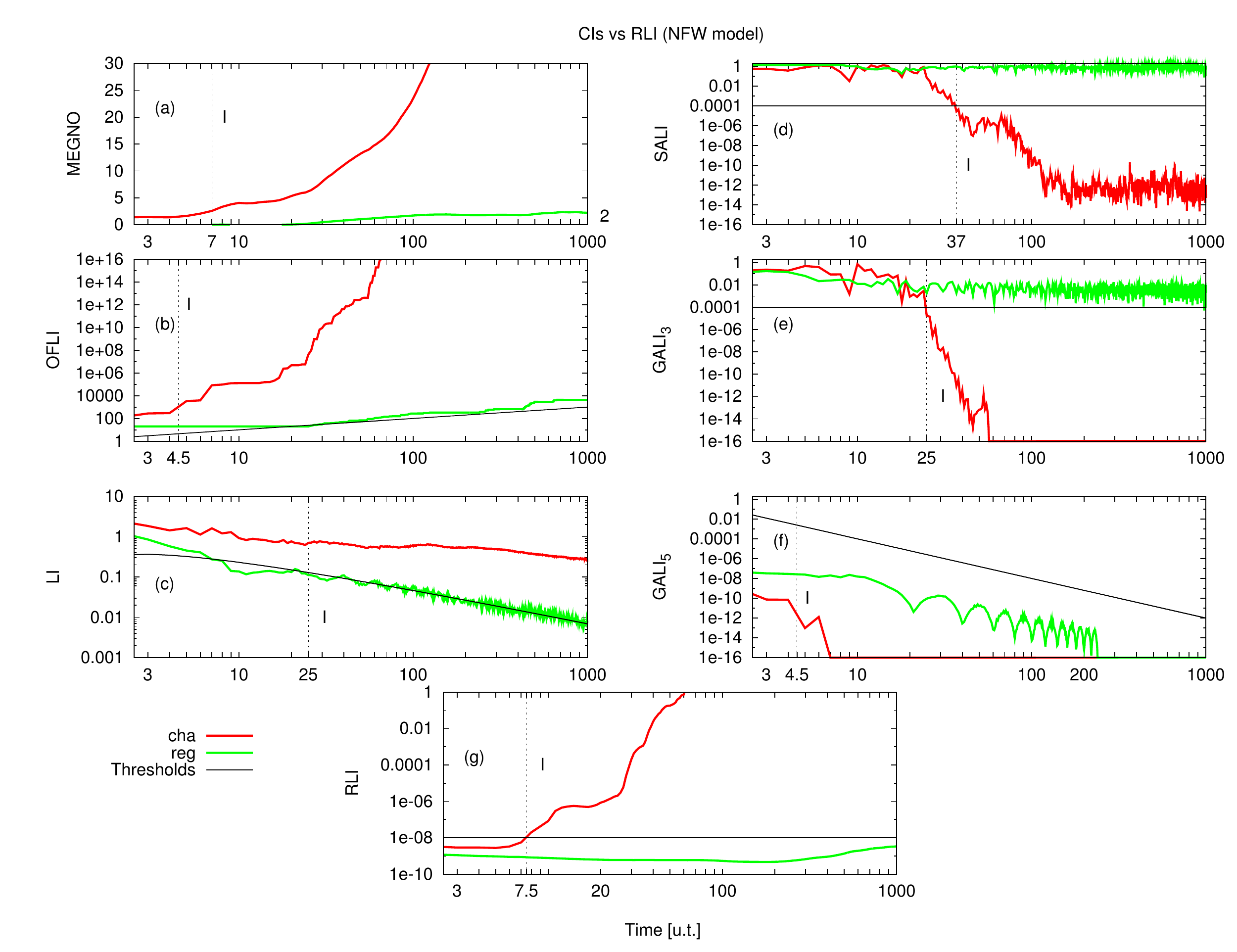}
\caption{Behaviours of (a) the MEGNO, (b) the OFLI, (c) the LI, (d) the SALI, (e) the GALI$_3$, (f) the GALI$_5$ and (g) the RLI for orbits ``cha'' and ``reg''. The thresholds as well as the line ``I'' are included (see text for further details).}
\label{nfw-regcao}
\end{figure}

In the next experiment, we follow two orbits in the NFW model for a time--span of 1000 Gyrs (i.e. $\sim77$ Hubble times) in order to have convergent final values of all the indicators in the study. We compute the time evolution curves of several indicators, including the RLI, and present the results for a chaotic and a regular orbit (``cha'' and ``reg'', respectively) in Fig. \ref{nfw-regcao}. In order to distinguish efficiently the chaotic orbit from the regular one, we proceed as in Sect. \ref{sec:3.2} and use the same thresholds used there for the MEGNO, the OFLI, the LI, the SALI and the RLI. The threshold for the GALI$_3$ in the NFW model will be the same constant used for the SALI (see \cite{CB2006}) because the model is a 3 degree of freedom (hereinafter d.o.f.) system. The threshold for the GALI$_5$ will be $t^{-4}$, with $t$ the time (see \cite{DMCG2012} for further details). In Fig. \ref{nfw-regcao}, we mark with the vertical line ``I'' the time after which the orbit ``cha'' is clearly identified as a chaotic orbit. 

Fig. \ref{nfw-regcao} shows that the accurate identification of the orbit ``cha'' as a chaotic orbit by the MEGNO (panel a), the OFLI (panel b), the GALI$_5$ (panel f) and the RLI (panel g) is made within a Hubble time ($\sim$13 Gyrs). The above mentioned indicators show that the orbit will behave as a chaotic orbit within a physical meaningful time--span (i.e. the age of the Universe) which is important to understand the dynamics of a real system like a galaxy.

In this section, we present results that support the RLI as a reliable indicator. This technique shows a phase space portrait very similar to those shown by the MEGNO, the OFLI or the GALI$_3$. Furthermore, the RLI identified the chaotic nature of the orbit ``cha'' very fast in the second experiment (the fastest being the OFLI and the GALI$_5$), within a Hubble time. These results put the RLI on an equal footing with the other fast variational indicators.

\subsection{A short discussion on the computing times}
\label{sec:3.4}

The computing times of the indicators are specially crucial for time--consuming simulations and their estimation helps for an efficient usage of the computational resources. However, such estimation is not an easy task. The computing times depend on a wide variety of factors such as the complexity of the model and that of the indicator's algorithm, its numerical implementation, the hardware, etc. Although making a detailed study of the computing times is not our concern (see \cite{DMCG2012} for further information on the subject), we would like to point out the fact that the easy algorithm of the RLI helps for a rather fast computation of the indicator. Furthermore, we are not dealing with the computing times themselves. We are going to register the ratios between the computing times of the different techniques and the computing time of the LIs\footnote{Here, the LIs are the numerical approximations of the spectra of Lyapunov Characteristic Exponents.} for orbits in two of the systems previously studied in the chapter, the HH and the NFW models. If the ratio is above 1, then the computing time of the indicator is larger than the computing time of the LIs. 

In this experiment, we used the following hardware/software configuration: an Intel Core i5 with four cores, CPU at 2.67 GHz, 3 GB of RAM, an OS of 32 bits, and the {\tt gfortran} compiler of {\tt gcc} version 4.4.4, without any optimizations. The code to compute the indicators is the \texttt{LP-VIcode}, the acronym for ``La Plata Variational Indicators code''. The alpha version of the \texttt{LP-VIcode} was introduced in \cite{DMCG2012a} and currently, it is a fully operational code that computes a suite of ten variational indicators (see \cite{CMD2014})\footnote{Further information on the \texttt{LP-VIcode} can be found at the following url: \texttt{http://www.fcaglp.unlp.edu.ar/LP-VIcode/}}. The initial setup of the \texttt{LP-VIcode} is the following: the integration time is 1000 u.t. (or 1000 Gyrs in the NFW model), the step of integration is 0.001 u.t. (or 1 Myr in the NFW model). 

We registered the ratios for two pairs of orbits and for every indicator. Both pairs have one chaotic and one regular orbit. One of the pairs of orbits is located in the HH model and the other in the NFW model. The computing time of the LIs for the HH model (i.e. 4 LIs) is 0m15.204s., and for the NFW model (i.e. 6 LIs) is 3m27.703s. The energy conservation is $\Delta E\sim10^{-13}$. 

The results are shown in Table \ref{timesregcao}. The value ``N'' is the number of d.o.f. of the system. For instance, in the HH model (second column in Table \ref{timesregcao}), we compute 4 LIs and 3 GALIs: the GALI$_2$, the GALI$_3$ and the GALI$_4$.  

\begin{table}\centering
\caption{Ratios of the computing times for several indicators, including the RLI, for the HH and the NFW models}
\label{timesregcao}
\begin{tabular}{p{3cm}p{3cm}p{3cm}}
\hline\noalign{\smallskip}
Indicator(s) & Ratios (HH model) & Ratios (NFW model) \\ 
\noalign{\smallskip}\svhline\noalign{\smallskip}
 (2N-1) GALIs & $\sim$2.026 & $\sim$1.024 \\
 (2N) LIs & 1 & 1 \\
 RLI & $\sim0.892$ & $\sim0.443$ \\
 SD & $\sim$0.613 & $\sim0.375$ \\
 SALI & $\sim$0.582 & $\sim$0.372 \\
 MEGNO & $\sim$0.53 & $\sim$0.174 \\
 FLI/OFLI & $\sim$0.432 & $\sim$0.151 \\
\noalign{\smallskip}\hline\noalign{\smallskip}
\end{tabular}
\end{table}

The RLI is not the least consuming indicator so far, according to the information given in Table \ref{timesregcao}. The FLI/OFLI, the MEGNO, the SALI and the SD might be more desirable options (in that order). Nevertheless, the easy implementation of the RLI helps to reduce significantly the computing time of the LIs, and its ratio is not much larger than those of the SD or the SALI.

\section{Application of the RLI to planetary systems}

The ongoing discovery of exoplanetary systems is certainly the most rapidly growing field of astronomy. Up to now the number of known planetary systems is almost 200. On the other hand, the masses and orbital elements of the planets detected can be determined only with uncertainties. Therefore it is of high importance to provide stability estimates of these systems. Before the launch of the space missions devoted to detect terrestrial planets (CoRoT, Kepler), only the massive giant planets were discovered mainly by the radial velocity method. One of the major applications of the RLI was to study the stability of the still hypothetical terrestrial planets in planetary systems containing at least one giant planet \cite{SSEPD2007}. Another possible application area of the RLI is to study the stability of different orbital solutions of resonant exoplanetary systems provided by radial velocity observations. In this review we shortly describe the stability studies done for the resonant planetary system HD 73526 \cite{SKK2007}. Finally, we present the applicability of the RLI to map the high order resonances in the restricted three-body problem, which might have relevance when studying the behavior of Kuiper belt objects \cite{ERSF2012}. We note that close relatives to our investigations are the works of \cite{V2008} computing the stability maps of the system 55 Cancri by using various indicators (LI, SALI and FLI), and of \cite{CKVH2007} studying the dynamical stability of the Kuiper-Belt using the LI indicator.

In all of the above problems the mean motion resonances (MMR) play an essential role, therefore we shortly summarize their properties. A MMR occurs between two bodies orbiting a more massive body if their orbital periods can approximately be expressed as a ratio of two positive integer numbers,
$T_1/T_2 = (p+q)/p$, where $T_1$ and $T_2$ denotes the orbital periods of the two bodies, respectively. A MMR can be characterized by studying the behaviour of the resonant angle, which in the model of the restricted three-body problem for an inner MMR is
\begin{equation}
\theta = (p+q)\lambda^\prime - p\lambda - q\varpi,
\label{innerres}
\end{equation}
while for an outer MMR can be written as
\begin{equation}
\theta = (p+q)\lambda - p\lambda^\prime - q\varpi,
\label{outerres}
\end{equation}
where $\lambda$ and $\varpi$ are the mean orbital longitude, and longitude of the pericenter of the massless body, while $\lambda^{\prime}$ is the mean orbital longitude of the massive body. If $\theta$ librates with a certain amplitude, the two bodies are engulfed in the $(p+q):p$ MMR.
In this way almost the same orbital configuration of the bodies involved in the given MMR is repeated. Depending on the relative positions of the bodies, this configuration can be protective, or can result in unstable orbits, see for more details in \cite{MD1999}.

\subsection{Stability catalogue of the habitable zones of exoplanetary systems}

The main idea behind the stability catalogue was to map the regions of a planetary system that can host dynamically stable terrestrial planets \cite{SSEPD2007}. The dynamical stability of a terrestrial planet is one of the strongest requirements for a habitable planetary climate. The most important requirement for the \emph{habitability} of a planet is to contain water in liquid phase on its surface. A region around a star, in which an Earth-mass planet could be habitable in the above sense is called as the \emph{habitable zone} (HZ), see \cite{KWR1993} and \cite{KRKetal2013} for more details.

\subsubsection{Used models and initial conditions}

The stability of terrestrial planets can be studied by using different approaches: (i) by detecting the stable and unstable regions of the parameter space of each exoplanetary system separately or (ii) by using stability maps computed in advance for a large set of orbital parameters. In the stability catalogue we presented such stability maps also showing how to apply them to the exoplanetary systems under study. This second approach has the advantage that the stability properties of a terrestrial planet can be easily reconsidered when the orbital parameters of the giant planet of an exoplanetary system are modified. This is very often the case, since the orbital parameters of the giant planets are quite uncertain, and due to the accumulation and improvement of the observational data, they are subject to change quite frequently. Instead of the re-exploration of the phase space of each individual exoplanetary system after possible modification of the orbital parameters of the giant planet, the stability properties of the investigated planetary system can be easily re-established from the already existing stability maps. These stability maps, which form a stability catalogue, can also be used to study the stability properties of the habitable zones of known exoplanetary systems. 

The majority of planetary systems which are detected so far consists of a star and a giant planet revolving in an eccentric orbit. Therefore, we used a simple dynamical model, the \emph{elliptic restricted three-body problem} in which there are two massive bodies (the primaries) moving in elliptic orbits about their common center of mass, and a third body of negligible mass moving under their gravitational influence (for details see \cite{Szebehelybook}). In our particular case the primaries are the star and the giant planet, and the third body is a small Earth-like planet, being regarded as massless. We note that among the extrasolar planetary systems there is a high rate of multiple planet systems. Thus, a more convenient model for the stability maps would be the restricted N-body problem (with N-2 giant planets, $N\geq4$). The presence of additional giant planets certainly enhance the instabilities induced by just one massive planet, turning the HZ of the system more unstable. The main source of instabilities are the \emph{mean motion resonances} (MMR) between the massive giant planet and the Earth-like planet. Thus, by mapping these resonances, we can find the possible regions of the instabilities in the HZs. On the other hand, the dynamical model with one giant planet also offers the most convenient way to display the most important MMRs as a function of the mass ratio of the star and the giant planet, and of the eccentricity of the giant planet. 

In the catalogue of dynamical stability one important quantity is the mass parameter of the problem $\mu = m_1/(m_0 + m_1)$, where $m_0$ is the stellar, and $m_1$ is the planetary mass. The mass parameter has been changed between broad limits ($10^{-4}-10^{-2}$) with various steps of $\Delta\mu$, in total the different stability maps have been calculated for 23 values of $\mu$. The giant planet was placed around the star in an elliptic orbit, with semi-major axis $a_1$, eccentricity $e_1$, argument of periastron $\omega_1$ and mean anomaly $M_1$. The semimajor axis $a_1$ was taken as unit distance $a_1 = 1$ during all simulations. The eccentricity $e_1$ was changed between 0.0 and 0.5 with a stepsize of $5\times 10^{-3}$. The argument of periastron was fixed at $\omega_1 = 0^\circ$, while the mean anomaly $M_1$ was changed between $0^\circ$ and $360^\circ$ with $\Delta M_1 = 45^\circ$.
The test planet was started in the orbital plane of the giant planet with an initial eccentricity $e = 0$, argument of periastron $\omega = 0^\circ$, and mean anomaly $M = 0^\circ$. The semi-major axis $a$ of the test planet was changed in two different intervals: (i) for orbits of the test planet `inside' the orbit of the giant planet between 0.1 and 0.9 with a stepsize of $\Delta a = 10^{-3}$ and (ii) for `outside' orbits between 1.1 and 4.0 with a stepsize of $\Delta a = 3.625\times 10^{-3}$. Further details and the complete catalogue can be found at \texttt{http://astro.elte.hu/exocatalogue/index.html}.

\subsubsection{Stability maps}

\begin{figure}[t]
\includegraphics[scale=0.55]{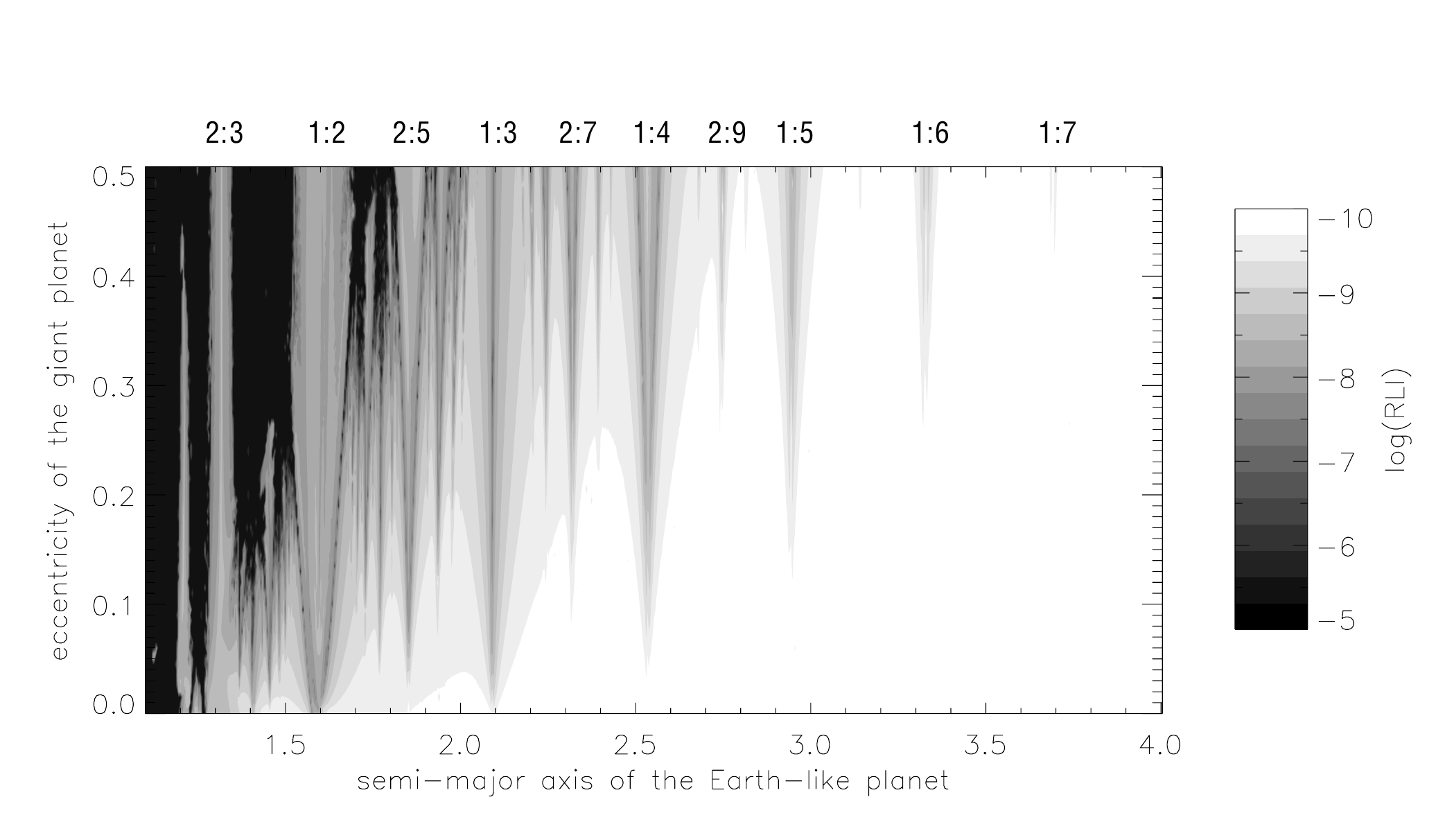}
\caption{Stability map of the outer MMRs for the Earth-like planets in the elliptic restricted three-body problem for $\mu = 0.001$ and $M=0^\circ$. White colour denotes ordered motion, light grey strips and ``V" shapes the different resonances, while black the strongly chaotic regions.}
\label{sandor1fig1}       
\end{figure}
\begin{figure}[t]
\includegraphics[scale=0.55]{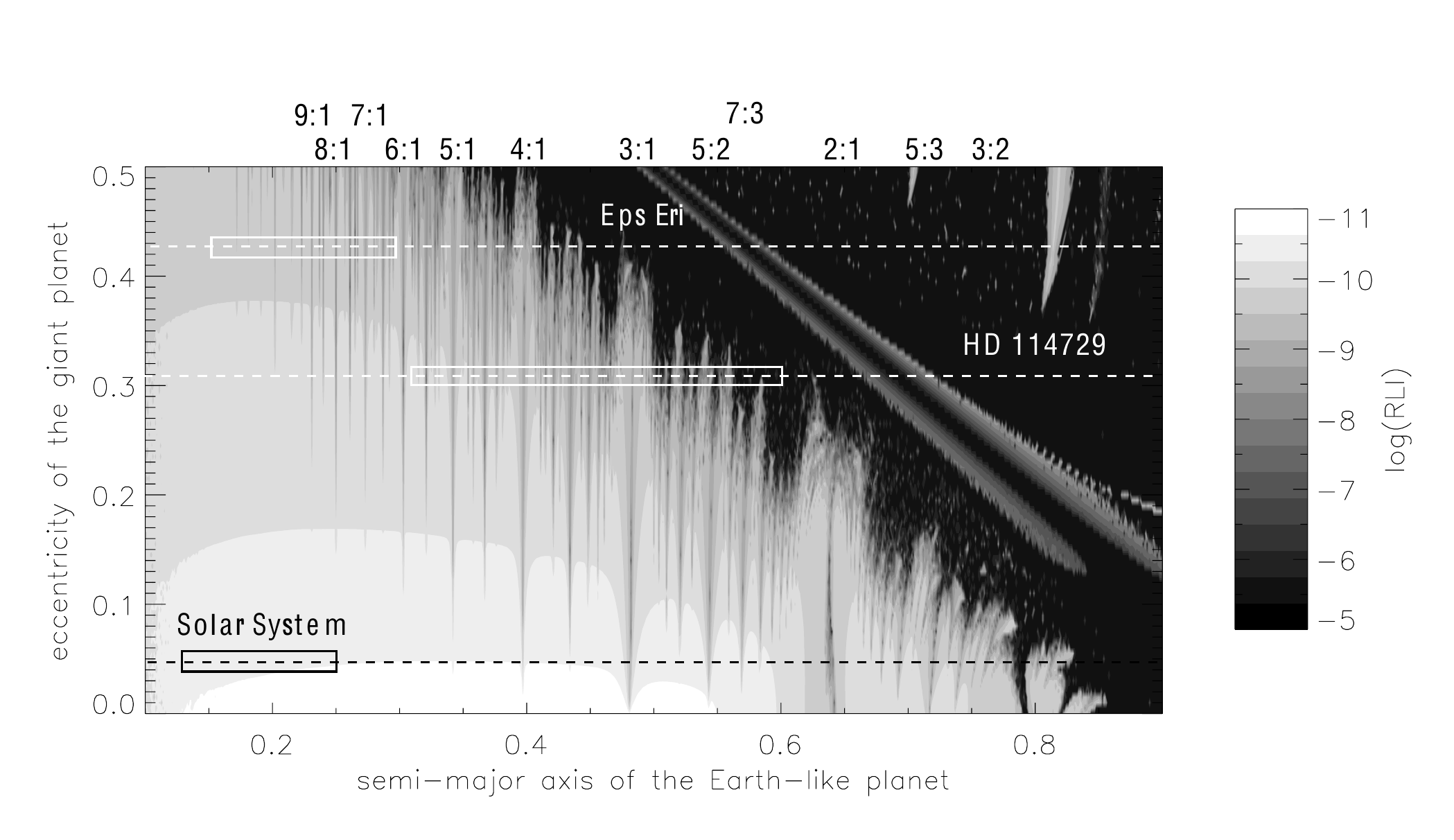}
\includegraphics[scale=0.55]{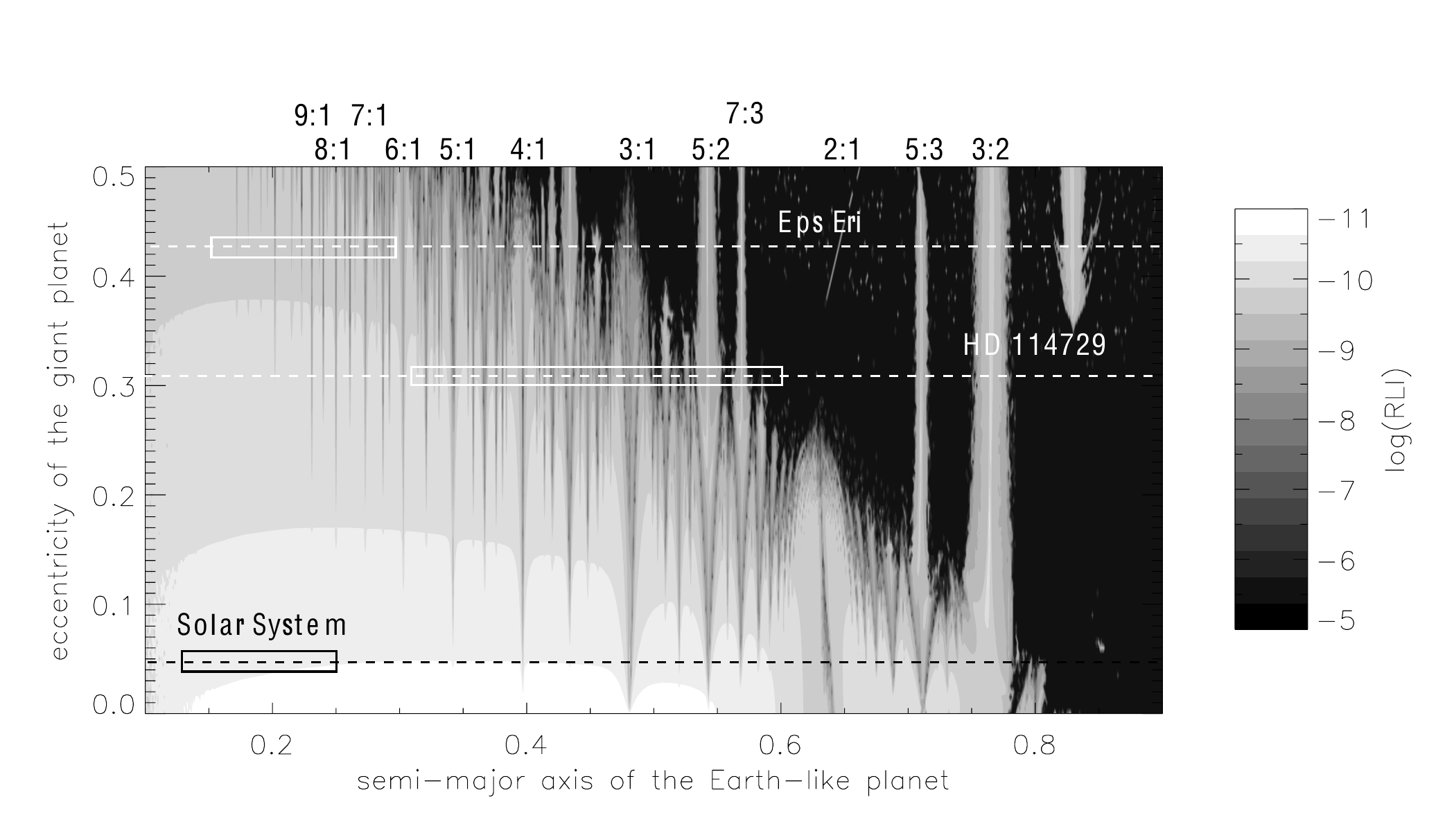}
\caption{Stability map of the inner MMRs for the Earth-like planets in the elliptic restricted three-body problem for $\mu = 0.001$, $M=0^\circ$ (upper panel), and $M=180^\circ$ (lower panel), respectively. We note the different character of the 5:2, 7:3, and 3:2 MMRs depending on the orbital positions between the test particle, and the perturbing body. White colour denotes ordered motion, light grey strips and ``V" shapes the different resonances, while black the strongly chaotic regions. We displayed the properly scaled HZs of three exoplanetary systems in the stability maps. Studying the figures one can conclude that the HZ of the Solar system is stable, the HZs of $\epsilon$ Eridani and HD 114729 are \emph{marginally stable} meaning that they are filled with several MMRs.}
\label{sandor1fig2}       
\end{figure}

Due to the very good visibility of the outer MMRs, we first display the case when the semi-major axis of the test planet is larger than the giant planet's semi-major axis ($a>1$). Additionally, in the stability map shown in Fig. \ref{sandor1fig1} the values $\mu = 0.001$ and $M=0^\circ$ were kept fixed. For each value of the RLI a gray shade has been assigned. White regions correspond to small RLI values, thus they are very stable. The MMRs appear either as light strips in the dark, strongly chaotic regions or as the well-known ``V"-shaped structures representing the separatrices between resonant and non-resonant motion. The inner regions of the resonances may be lighter than the lines of the bounding separatrices indicating regular motion in a protective resonance (e.g. the 2:5 MMR). Near the separatrices the motion is always chaotic, moreover at some MMRs even the inner part of the resonance is chaotic as indicated in the case of the 1:3 MMR, for instance. By increasing the giant planet's eccentricity $e_1$ many resonances overlap giving rise to strongly chaotic and thus very unstable behaviour. The reason of this phenomenon is that by increasing $e_1$ the apocenter distance of the giant planet also increases, therefore the giant planet perturbs more strongly the outer test planet.

In stability maps displayed in Fig. \ref{sandor1fig2} the semi-major axis of the test planet is smaller than that of the giant planet ($a<1$). The two panels for the mass parameter $\mu=0.001$ show the stability maps for the two different starting positions of the giant planet, $M_1 = 0^\circ$ (upper panel) and $M_1 = 180^\circ$ (lower panel), respectively. Between the test planet and the giant planet, a large number of inner MMRs can be found, which dominate the stability maps. Inside the resonances the stable or chaotic behaviour of the test planet depends on the initial angular positions of the two planets. This is clearly visible by comparing the two panels. On the other hand the location of the MMRs is not altered, since this depends on the ratio of the semi-major axes of the two planets. In the lower panel of Fig. \ref{sandor1fig2} several MMRs (5:2, 5:3, 3:2) are stable, which is not the case in the upper panel of Fig. \ref{sandor1fig2} This is due to the fact that the relative initial positions determine the places of conjunctions of the two planets. If they meet regularly near the pericenter of the giant planet, the motion of the test planet becomes chaotic, while it can remain regular if the conjunctions take place near the apocenter of the giant planet. The effect of the initial phase difference between the planets is important, therefore a bunch of stability maps have been prepared for more initial values of the mean anomaly $M_1$ of the giant planet. These stability maps can be found in the online exocatalogue (\texttt{http://astro.elte.hu/exocatalogue/index.html}). By increasing the value of the mass parameter $\mu$ it becomes clearly visible that the larger mass of the giant planet results in stronger perturbations, and therefore more enhanced chaotic region.

\subsubsection{Stability of terrestrial planets in the habitable zones}

\begin{figure}[h]
\includegraphics[scale=0.55]{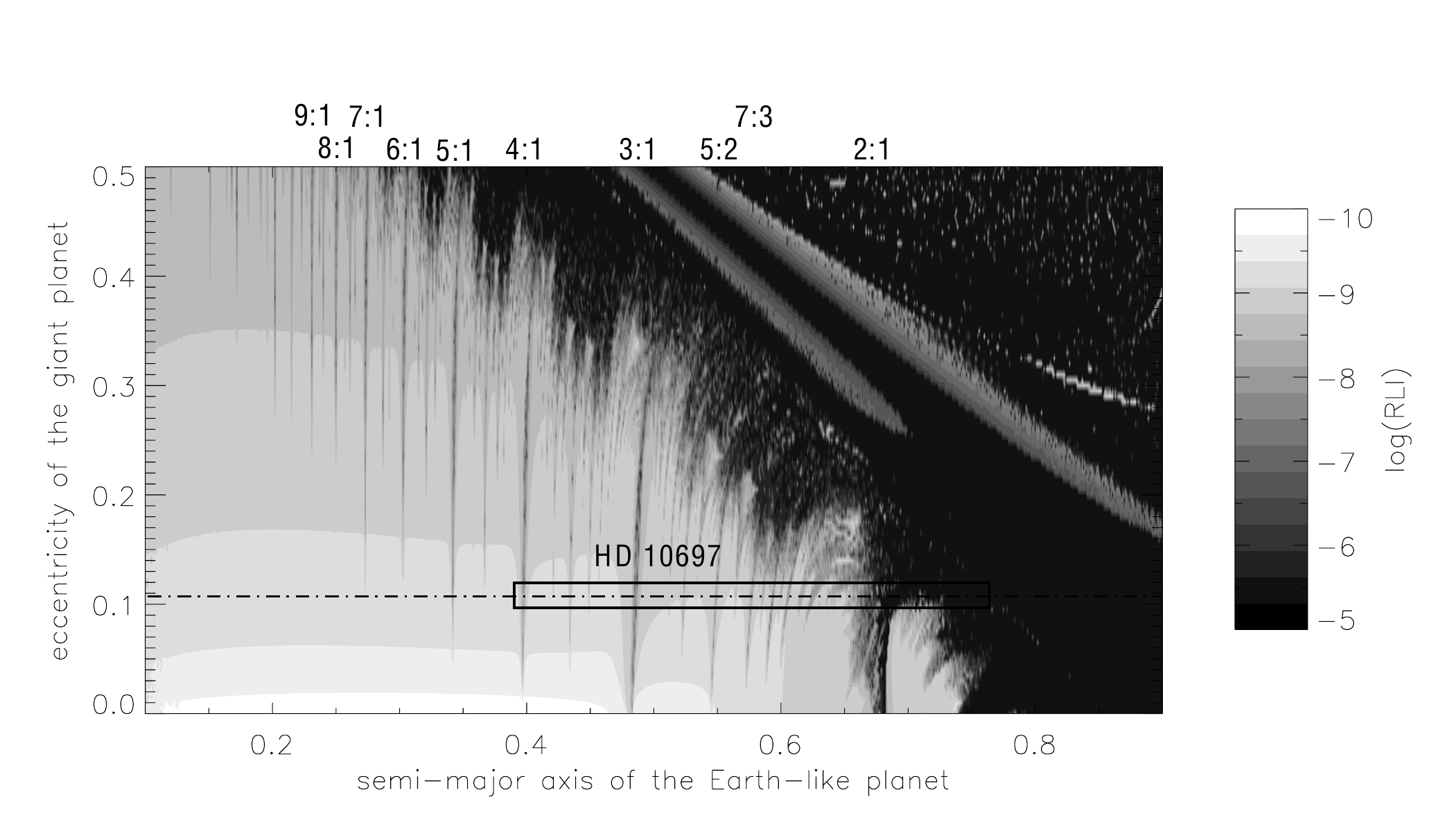}
\caption{Stability map for inner orbits of an Earth-like planet, when $\mu=0.005$ and $M_1 =0^\circ$. The line $e_1 =0.11$, corresponds to the system HD 10697. The scaled HZ of this system is between $0.39<a<0.77$. Its inner part is stable (containing only a few weakly chaotic MMRs), while the outer part is in the strongly chaotic regions.}
\label{sandor1fig4}       
\end{figure}

In this section, we show how to use the catalogue to determine the stability of hypothetical Earth-like planets in exoplanetary systems. As an example, we consider the case of HD10697, where $a_1 = 2.13$ AU, $e_1 = 0.11$ and $\mu=0.0055$. Fig. \ref{sandor1fig4} shows a stability map, calculated for $\mu = 0.005$ for inner orbits of the test planet. This corresponds to the minimum mass of the giant planet (minimum masses are used throughout). The stability of a small planet (starting with $e = 0$) in the system HD10697 can be studied along the line $e_1 = 0.11$. One can see that for small semimajor axes, $a<0.33 a_1=0.729$ AU the parameter space is very stable. When $a>0.33 a_1$ several resonances appear, among which the most important are the 5:1, 4:1, 3:1 and 2:1 MMRs. For $a>0.73 a_1=1.55$ AU, a strongly chaotic region appears. The classical HZ of this system is between 0.85 and 1.65 AU therefore in Fig. \ref{sandor1fig4} the scaled classical HZ is located between $0.85/a_1 = 0.39$ and $1.65/a_1 = 0.77$ (shown as a rectangle, elongated in horizontal direction). One can see that the inner part of the classical HZ contains ordered regions, but stripes of certain resonances are also present. The outer part of the classical HZ is in the strongly chaotic region.

\subsection{Stability of resonant exoplanetary systems}

A significant amount of multiple extrasolar planetary systems contain pairs of giant planets which are orbiting in MMRs. The in situ formation of resonant planetary systems is very unlikely, since each resonance requires certain ratio of the semi-major axes. The more favorable scenario is the \emph{type II migration} of giant planets being embedded in the still gas rich protoplanetary disc. Type II migration appears when a massive giant planet carves a gap in the gaseous protoplanetary disc practically inhibiting the gas flow through the gap. In that case the planet's semi-major axis is changing according to the viscous evolution of the protoplanetary disc, see more about the topic in (\cite{BM2013}).

If the migration of two giant planets is convergent (e.g. the difference between their semi-major axes is decreasing) the phenomenon of the \emph{resonant capture} will occur between them, and the two planets can migrate very close to their host star. The efficiency of migration is excellently demonstrated by hydrodynamic simulations modelling the formation of the resonant system around the star GJ 876 \cite{KLMP2005}. There are other planetary systems in which the giant planets reached the 2:1 MMR through type II migration such as HD 128311 \cite{SK2006}, and HD 73526 \cite{SKK2007}. 

\begin{figure}[t]
\sidecaption[t]
\begin{tabular}{cc}
\includegraphics[scale=0.7]{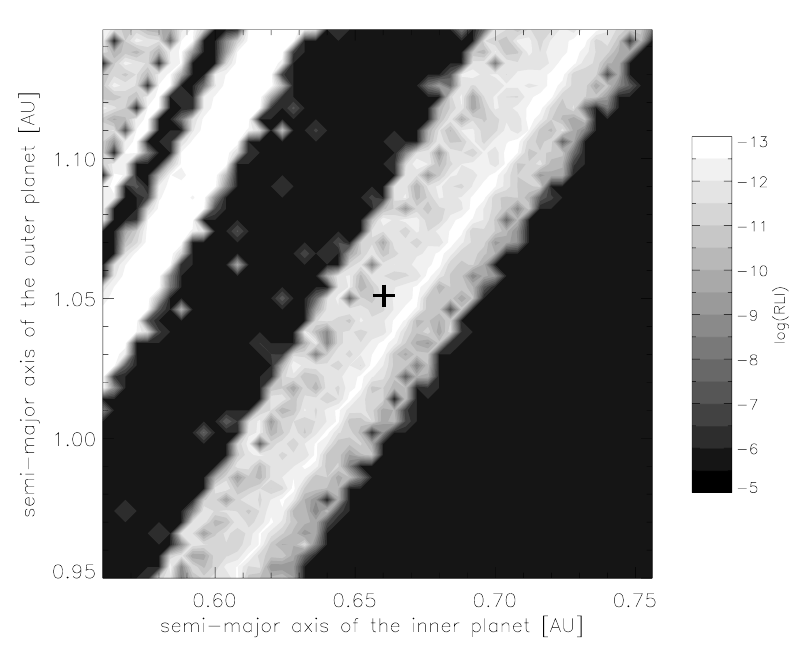}&
\includegraphics[scale=0.7]{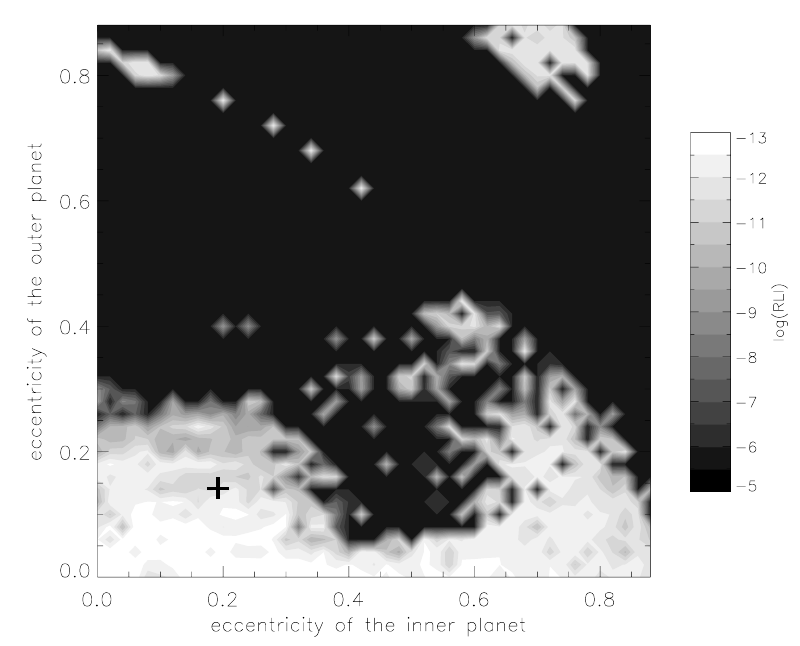}
\end{tabular}
\caption{Stability maps calculated around the orbital elements of \cite{TBMetal2006}. The structure of the stability maps indicates that the orbital elements (marked by “+”) are in a weakly chaotic region. Here white colour refers to ordered, lighter grey shades to weakly chaotic, and darker shades to unstable regions.} 
\label{sandor2fig12}
\end{figure}

\begin{figure}[t]
\sidecaption[t]
\begin{tabular}{cc}
\includegraphics[scale=0.7]{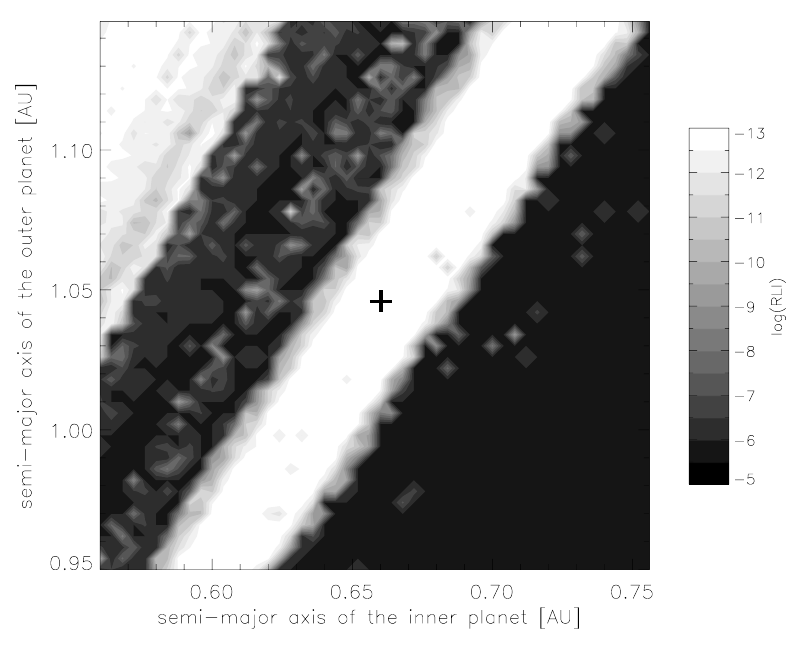}&
\includegraphics[scale=0.7]{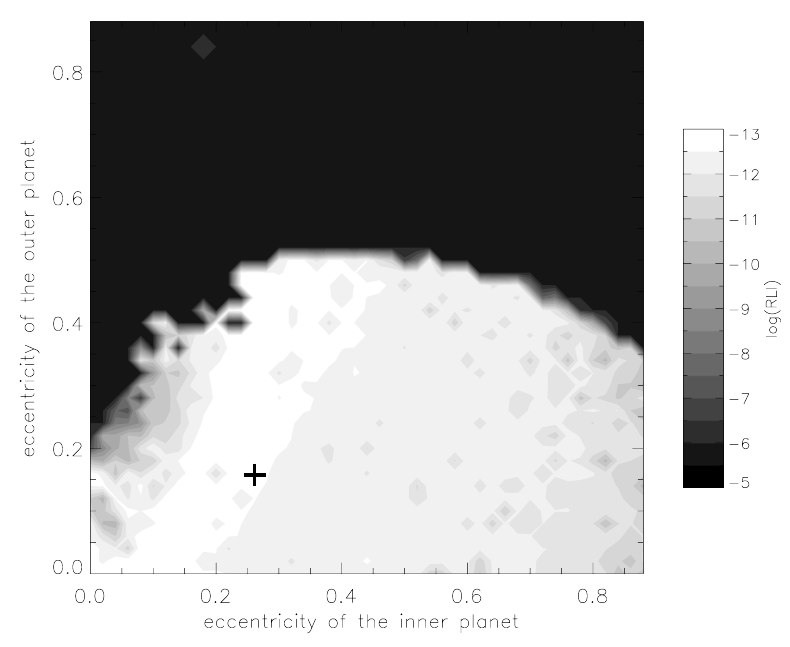}
\end{tabular}
\caption{Stability maps calculated around one set of orbital elements given in \cite{SKK2007} It can be seen from the stability maps that the orbital elements (marked by “+”) are embedded well in the stable region marked by white colour.} 
\label{sandor2fig34}
\end{figure}

The detection of giant planets is based on the radial velocity method. To calculate the orbital elements of the planets of a multiple system is not an easy task. Although there are well-known and widely used algorithms to provide reliable orbital fits, the orbital elements obtained not always result in a stable configuration for the planetary system. Regarding the resonant system of giant planets around HD 73526, \cite{TBMetal2006} published orbital elements and planetary/stellar masses which resulted in \emph{stable} orbits of the giant planets over 1 million years. On the other hand, the orbits are \emph{chaotic}, as was clearly visible from numerical integrations of the three-body problem using the given elements as initial conditions. Since chaotic behavior may be uncommon among the resonant extrasolar planetary systems, and may not guarantee the stability of the giant planets for the whole lifetime of the system (being certainly longer than 1 million years), we searched for regular orbital solutions for the giant planets as well \cite{SKK2007}. As a first attempt to study the degree of the chaoticity, we mapped the parameter space around the solution of \cite{TBMetal2006}. We have calculated the stability properties of the $a_1-a_2$, $e_1-e_2$, $M_1-M_2$, and $\varpi_1-\varpi_2$ parameter planes, where $a$ is the semi-major axis, $e$ is the eccentricity, $M$ is the mean anomaly and $\varpi$ is the longitude of periastron of one of the giant planets. 

In Fig. \ref{sandor2fig12} the stability structures of the parameter planes for the semi-major axis and the eccentricities are displayed. During the calculation of a particular parameter plane the other orbital data have been kept fixed to their original values. On each parameter plane the stable regions are displayed by white, the weakly chaotic regions by grey, and the strongly chaotic regions by black colors. The values of the corresponding orbital data are marked on each parameter plane. By studying the stability maps, one can see that the orbital elements given by \cite{TBMetal2006} are located in a weakly chaotic region, which explains the irregular behavior of the planetary eccentricities. We again stress that this does not automatically imply the instability of their fit, however by using these orbital data the system yields chaotic behavior and can be destabilized in longer timescales. Studying the stability maps it can also be conlcuded that the fit cannot be easily improved by the simple change of one of the orbital elements. The parameter plane is almost entirely weakly chaotic, there exists only a narrow strip of ordered behavior. After obtaining completely new sets of orbital elements using the Systemic Console (\cite{MSWetal2009}) the stability maps around these fits were recalculated. In Fig. \ref{sandor2fig34} parameter planes for the semi-major axis and the eccentricities are displayed, now around one of the stable orbital solutions. It can be clearly seen that the new orbital solution is well inside the region for ordered motion, which provides stability for the whole lifetime of the system. Based on the above example, it can be concluded that the RLI performs well in stability investigations of extrasolar planetary systems. 

\subsection{Application of the RLI to study libration inside high order MMRs}

The most recent application of the RLI in dynamical astronomy has been presented by \cite{ERSF2012} investigating the dynamical structure of high order resonances in the elliptic restricted three-body problem. This study has relevance in dynamics of Kuiper-belt objects. Moreover, it proved through  numerical integration of a large set of orbits that the RLI excellently indicates the libration of the resonant angle inside a MMR. In what follows, first we present the results of \cite{ERSF2012} obtained for the 3rd order $8:5$ inner resonance, in which case according to Eq. (\ref{innerres}) the resonant variable is
\begin{equation}
\theta = 8\lambda^{\prime} - 5\lambda - 3\varpi,
\end{equation} 
where $\lambda$ and $\varpi$ are the mean orbital longitude, and longitude of the pericenter of the inner body, while $\lambda^{\prime}$ is the mean orbital longitude of the outer body. 
 
\begin{figure}[t]
\includegraphics[scale=0.95]{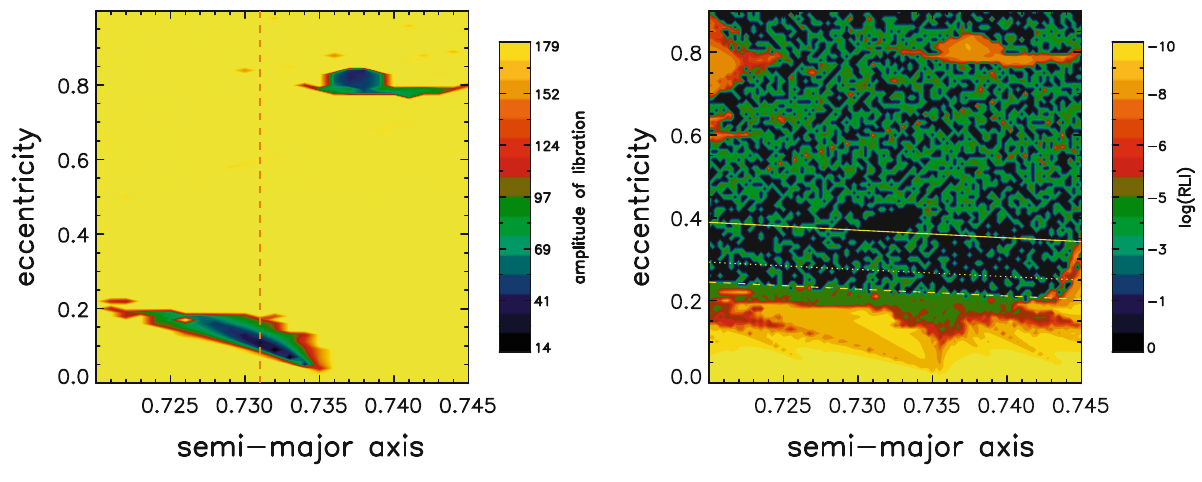}
\caption{The 8:5 inner MMR (color online). Left panel: the two regions of libration as displayed by the libration amplitude of the resonant variable. The exact location of the MMR is at $a_r = 0.7310$ in normalized units. Right panel: the RLI map around the resonance, in which the two regions of libration can be clearly seen. The continuous curve denotes the unit apocentre distance, while the dotted and dashed curves the unit apocentre distance decreased by 1, and 1.5 Hill’s radius, respectively.}
\label{erdifig1}       
\end{figure} 
 
A simple, but computationally demanding way to map the neighbourhood a MMR on the $a-e$ plane is to calculate the libration amplitude of $\theta$ for orbits whose initial semi-major axis and eccentricity values are taken from a grid, and the test particle has been started from pericentre. Such calculation can be seen on the left panel of Fig. \ref{erdifig1}. Using the same grid resolution and initial conditions the RLI values were also calculated, see the right panel of Fig. \ref{erdifig1}. The exact location of the resonance (when $\theta = 0^\circ$) is marked with the vertical dashed line in the left panel of Fig. \ref{erdifig1}. The colour code corresponds to the value of the libration's amplitude: the darker the colour, the smaller the amplitude. 

The $8:5$ MMR has an interesting structure, there is a libration for low values of the eccentricities $e<0.2$, and also for very high values $0.75<e<0.85$. Although the picture obtained by the RLI is more detailed, also showing some other neighbour MMRs, gives back the region of libration very well.  

The RLI has also been applied to study the high order outer MMRs, in which case the massless body's orbit is outside the massive planet's orbit. In the study of \cite{ERSF2012} the outer resonances of the Sun-Neptune system have been studied in two different models. These are the restricted three-body problem, in which only the gravitational effects of the Sun and Neptune were included, and also in a model in which the four giant planets were also taken into account. We will present here the results obtained for two different MMRs, namely the 8:5 and 7:3 outer resonances in the model of the restricted three-body problem. We note that the resonant variable of an outer $(p+q):p$ MMR is given by Eq. (\ref{outerres}).

\begin{figure}[t]
\includegraphics[scale=0.95]{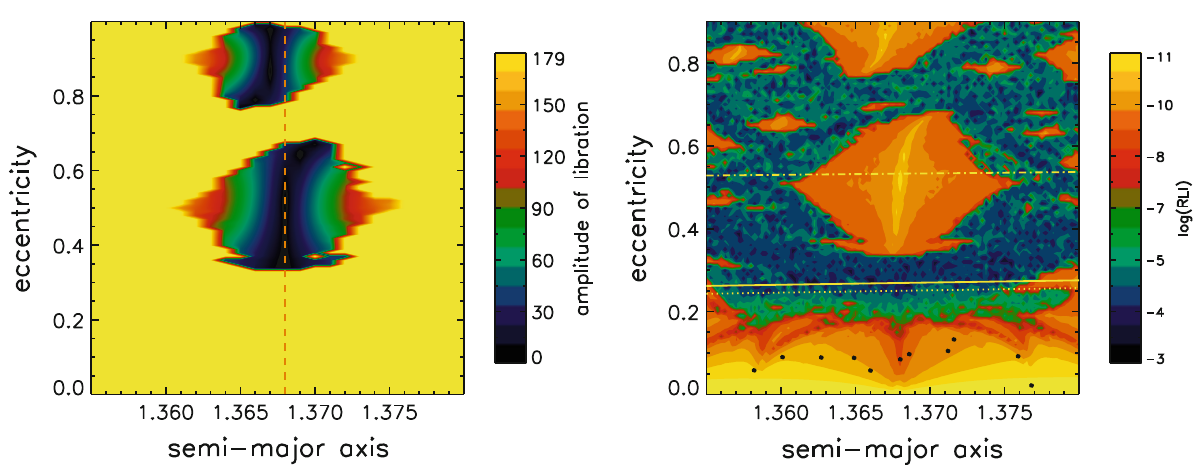}
\caption{The 8:5 outer MMR (color online). Left panel: the two regions of libration displayed by the libration amplitude of the resonant variable. The exact location of the MMR is at $a_r = 1.3680$ in normalized units. Right panel: the RLI map around the resonance, in which the regions of libration are clearly visible. The continuous curve indicates the unit pericentre distance, while the dotted curve the unit pericentre distance increased by one Hill’s radius. The dash-dotted curve shows the place of Uranus-crossing. The filled circles mark circulating TNOs.}
\label{erdifig2}       
\end{figure} 

Similarly to the $8:5$ inner MMR, the neighbourhood of the exact $8:5$ outer MMR ($a_r = 1.3680$ in normalized units) has been mapped on a dense grid of the $a-e$ plane (the test particle has been started from apocentre) and the libration regions are marked with different shades corresponding to the amplitude of $\theta$. There are two regions of libration in this resonance located at relatively high eccentricities ($0.35<e<0.7$, $0.75<e$), see the left panel of Fig. \ref{erdifig2}. Using the same $a-e$ grid the RLI values have also been calculated, see the right panel of Fig. \ref{erdifig2}. At the location of the exact resonance there is a nearly vertical dark strip indicating weak chaotic behaviour, and also the fact that in this configuration the $8:5$ MMR is not protective for low values of the test particle's eccentricity. The origin of the strong chaotic behaviour is due to the crossing of the orbits of the test particle and the perturbing body (Neptune in this case). There are different lines plotted to the RLI map. The continuous curve is the pericentre distance, the dotted curve is the pericentre distance increased by the Hill's radius of Neptune, and finally, the dash-dotted curve is the place of Uranus crossing. This latter curve may indicate that the high eccentricity libration regions of Fig. \ref{erdifig2} being present in the model of the restricted three-body problem, might vanish in more advanced models including the planet Uranus, for instance. Finally, we roughly compare the $8:5$ outer resonance with the behaviour of some of the existing TNOs having inclination $i<25^\circ$, see the black filled circles for the corresponding $a$ and $e$ values of these objects. All of the TNOs displayed are circulating, and really they occupy the low eccentricity regions of the $a-e$ plane. We note, however, that the $a$ and $e$ values of the TNOs have different epochs, so their positions does not reflect the actual state of the system. On the other hand, it can be clearly seen in Fig. \ref{erdifig2} that the TNOs at the $8:5$ MMR have circulating resonance variable.

\begin{figure}[t]
\includegraphics[scale=0.95]{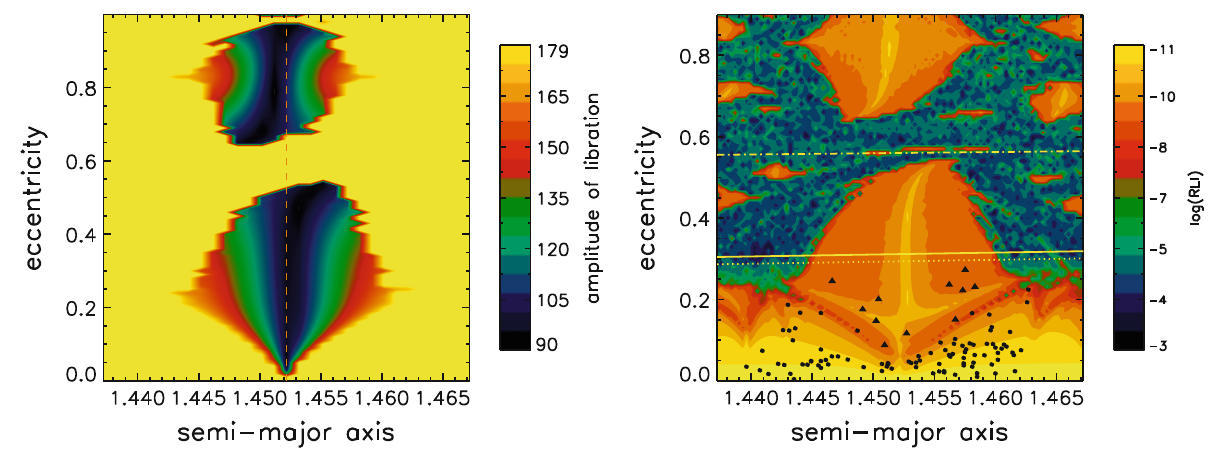}
\caption{The 7:4 MMR (color online). Left panel: the two regions of libration displayed by the libration of the resonant variable. The exact location of the resonance is at $a_r = 1.4522$. Right panel: the RLI map around the resonance. The triangles mark librating, while the filled circles circulating TNOs. For the description of the curves see the previous figures.}
\label{erdifig3}       
\end{figure}

In order to have a more complete picture of the high order outer resonances, we also summarize the case when a MMR has a protective character, and the resonant variable of bodies lying in its vicinity can both librate and circulate. A good candidate for this purpose is the $7:4$ 3rd order outer MMR. The exact resonance is at $a_r = 1.4522$ (normalized units). Similarly to the $8:5$ outer MMR, this resonance also has two regions of libration, but in this case the lower region of libration allows libration of test particles having low eccentricities, see Fig. \ref{erdifig3}. The right panel shows the dynamical structure of the resonance, and also TNOs found at this resonance. Most of them have circulating $\theta$, but there are a few of them in the librating region, too. Studying the right panel of Fig. \ref{erdifig3}, one can see that the librating TNOs are clearly below the Neptune-crossing line, while the region of libration extends above it.

As an overall conclusion we can state that the RLI is a very reliable tool in detecting the positions of the lower and higher order MMRs being present in various planetary systems. 

\section{Discussion and summary}
In this chapter we summarize the basic properties of the recently introduced chaos detection method, the RLI. The RLI is based on the time evolution of the infinitesimally small tangent vector to the orbit, which is provided by solving numerically the variational equations. Hence, the RLI belongs to the family of the so--called variational indicators. Although the definition of the RLI is based on that of the LI, in this review we give evidence that the distinction between regular and chaotic motion is much clearer with the RLI, which makes it a more reliable alternative than the LI. 

According to the comparative study with some wide--spread variational indicators, the RLI shows convincing performances in the experiments and considerably improves the performances of the classical LI. In generality, indicators like the FLI/OFLI or the MEGNO (actually there is a strong relationship between both indicators, see \cite{MCG2011} for further details) are usually believed to be better options for a general analysis of the structure of the phase space. Therefore, our study reinforces the fact that the RLI can also be used as an alternative technique, which operates with reasonable computing times to make conclusive pictures of the dynamics despite the complexity of the problem.

Based on the comparative work presented in Section \ref{sec:3}, we can summarize both the advantages and disadvantages of the RLI. In what follows we list its favorable properties/advantages, and also add that in which dynamical systems have been done the corresponding simulations.

\begin{itemize}
\item Having determined a reliable threshold in the H\'enon--Heiles model, the RLI (and the FLI/OFLI) shows the best approximation rates in the ordered/chaotic regions. We note that the methods SALI and GALI also estimate the true percentage of the ordered/chaotic orbits but with a slight slower way.
\item Comparing to the other chaos indicators also in the H\'enon--Heiles model, the RLI detects much faster the orbits from the large chaotic sea (e.g. the ``c-cs" orbit), than the other indicators. 
\item Studying a 4D symplectic mapping the RLI have been compared to the indicator MEGNO(2,0). In this case both the RLI and the MEGNO(2,0) reveal the fine structure of the phase space very accurately (much better than the LI, for instance). As a result of this experiment we also conclude that the RLI is a very effective tool in the characterization of a large array of initial conditions. 
\item In a complex 3D potential resembling a triaxial galactic halo (the so called NFW model), the RLI (together with the MEGNO, the OFLI, and the GALI$_5$) identifies the chaotic orbits within a Hubble time ($\sim$ 13 Gyrs). These indicators show that chaotic orbits can be identified within a physically meaningful time (i.e. the age of the Universe), which is important when studying the dynamics of a galaxy.
\end{itemize}

\noindent
On the other hand, when applying the RLI one should be aware that:
\begin{itemize}
\item Among the presented CIs, in the H\'enon--Heiles model the RLI identifies the so called ``c-sl" orbit in the slowest way. 
\item In the study performed in the 4D symplectic mapping, the RLI cannot really distinguish between chaotic and sticky orbits. This is a disadvantage if one is interested in detecting the sticky orbits. On the other hand, if we are interested in detecting \emph{all} chaotic orbits (including sticky orbits, as well), the application of the RLI might be useful. 
\item Regarding the time needed to calculate the RLI, we can conclude that it is not the least time consuming indicator. The FLI/OFLI, the MEGNO, the SALI and the SD might be more desirable options if the computation time really matters. We note, however, that with the current generation of fast computers this option became less important.
\end{itemize}

\noindent
In the last section of the current work we summarize the application of the RLI to planetary systems, which is its major application area. These studies include the development of a stability catalogue of hypothetical terrestrial exoplanets in extrasolar planetary systems, stability studies of resonant planetary systems and the investigation of high order mean motion resonances having relevance in studying the dynamics of the Kuiper--belt objects. We find that the RLI is an efficient and reliable numerical tool to map and characterize the dynamical structure of various mean motion resonances.

We note that the RLI (together with the SALI) has also been applied to map the stability regions of the Caledonian symmetric four--body problem, (\cite{SESS2004}). Since the preceeding studies are mainly related to detecting the chaotic behaviour occured near to resonances, the study on the Caledonian restricted four--body problem does not really belong to this line, thus to keep the lenght of the present study tractable, we omited its presentation here.

We would also like to remark that the very simple computation of the RLI from the widespread well--known LI and its better performances reported in very different scenarios, make the RLI a serious candidate to replace the LI in a variety of fields, and not only in dynamical astronomy. For instance, in a paper published in a journal of Chemical Physics, the RLI is used as the default chaos indicator in the Lyapunov weighted path ensemble method. One of the capabilities of the method is to identify pathways connecting stable states which are relevant in the context of activated chemical reactions (see \cite{GD2010}). 

Finally, the reliability of the RLI as a chaos indicator has been strongly demonstrated throughout this study, and as a result, the choice of the RLI to analyze a general dynamical system is well--founded.

\begin{acknowledgement}
The authors thank the invitation and support of the scientific coordinators of the international workshop on Methods of Chaos Detection and Predictability: Theory and Applications: Georg Gottwald, Jacques Laskar and Haris Skokos and the hospitality of the Max Planck Institute for the Physics of Complex Systems where the meeting took place. ZsS is supported by the J\'anos Bolyai Research Scholarship of the Hungarian Academy of Sciences. NM is supported with grants from the Consejo Nacional de Investigaciones Cient\'ificas y T\'ecnicas de la Rep\'ublica Argentina (CCT - La Plata) and the Universidad Nacional de La Plata.
\end{acknowledgement}

%
%
%

\end{document}